\pdfoutput=1
\documentclass[11pt,a4paper]{article}
\usepackage[margin=1in]{geometry}
\usepackage[T1]{fontenc}
\usepackage{lmodern}
\usepackage{natbib}
\usepackage{graphicx}
\usepackage{authblk}
\usepackage{microtype}
\usepackage{setspace}
\usepackage{amsfonts,amssymb,graphics,verbatim,bm,latexsym,amsmath,url,amsbsy}

\newtheorem{theorem}{Theorem}
\newtheorem{assumption}{Assumption}
\newtheorem{proposition}{Proposition}
\newtheorem{corollary}{Corollary}
\newtheorem{lemma}{Lemma}

\newtheorem{remark}{Remark}

\renewcommand{\thesection}{\arabic{section}}
\renewcommand{\theequation}{\arabic{section}.\arabic{equation}}

\renewcommand{\thelemma}{\arabic{section}.\arabic{lemma}}

\newcounter{bean}

\title{The inclusive synthetic control method}
\author[1]{Roberta Di Stefano}
\author[2]{Giovanni Mellace}
\affil[1]{University of Molise, Campobasso, Italy\\ \texttt{roberta.distefano@unimol.it}}
\affil[2]{University of Southern Denmark, Odense, Denmark\\ \texttt{giome@sam.sdu.dk}}
\date{October 2026}

\begin{document}
\maketitle

\begin{abstract}
We introduce the inclusive synthetic control method (iSCM), an extension of the traditional synthetic control framework that allows units potentially affected by an intervention—either directly or through spillovers—to remain in the donor pool. Because synthetic control weights are estimated using only pre-intervention data, constructing a synthetic control for both the main treated unit and each potentially affected unit yields a linear system that links the contaminated estimates to the true effects of interest. Solving this system exactly removes the contamination without altering the underlying estimator. We illustrate the empirical relevance of our iSCM by re-estimating the effect of German reunification on GDP per capita, allowing for spillovers from West Germany to Austria.


\end{abstract}

\section{Introduction}
\setcounter{equation}{0}

The synthetic control method (SCM), introduced by \citet{Aba2003} and further developed by \citet{Aba2010} and \citet{Aba2015}, was originally designed to estimate the causal effect of a policy intervention in settings with a small number of treated and control units observed over an extended time period, and has since been adapted to a much wider range of applications. SCM constructs a weighted average of control units, the synthetic control, that mimics what would have happened to the treated unit in the absence of the intervention. The weights are chosen to minimize the discrepancy between the pre-intervention characteristics of the treated unit and those of the synthetic control, and the causal effect is estimated as the post-intervention difference between the observed outcome of the treated unit and that of its synthetic control.

A key assumption of SCM is that the control group, often called the donor pool, contains only units that are unaffected by the intervention. This requirement is problematic in at least two scenarios: (a) some treated units would ideally be included in the donor pool to improve the pre-intervention fit; and (b) some control units are indirectly affected by the intervention, but excluding them from the donor pool substantially deteriorates the pre-intervention fit. When such \textit{potentially affected} units are included in the donor pool without correction, standard SCM estimators are contaminated by the post-intervention effects on these units. The natural alternative, restricting the donor pool to unaffected units, avoids this contamination but may come at the cost of a much poorer pre-intervention fit. This trade-off calls for a procedure that exploits the pre-intervention information contained in potentially affected units while still delivering a counterfactual based on uncontaminated outcomes.

As a motivating example, consider the German reunification study of \citet{Aba2015}, who suggest that reunification may have had spillover effects on neighbouring countries, in particular Austria; see also \citet{Aba2020}. Austria receives the largest weight (0.42) in the construction of synthetic West Germany, so even a moderate spillover effect could induce a sizeable bias. At the same time, excluding Austria from the donor pool substantially worsens the quality of the pre-intervention match and produces estimates of the reunification effect that are on average about 42\%  more negative than those obtained with the method we propose; the same pattern arises with the penalized SCM of \citet{Aba2019} (see Section~\ref{sec:application}). Moreover, excluding West Germany from Austria's donor pool when estimating the spillover effect yields implausibly positive estimates of the effect of reunification on Austria's gross domestic product (GDP).

Our main contribution is the inclusive synthetic control method (iSCM), a procedure that removes the post-intervention effects of the intervention from the outcomes of potentially affected donor units, so that these units can safely remain in the donor pool. The method exploits the fact that synthetic control weights are estimated using pre-intervention data only, a period in which no unit is affected. When potentially affected units are kept in the donor pool, the synthetic control of the main treated unit is a weighted average of donor outcomes that absorbs the effects of the intervention on those units. The resulting estimate therefore equals the true effect minus a known linear combination of the effects on the potentially affected units. By constructing a synthetic control for every potentially affected unit as well, including the main treated unit in the respective donor pools, we obtain a system of $m$ linear equations in $m$ unknowns, namely the effect on the main treated unit and the $m-1$ effects on the potentially affected units. Solving this system delivers our estimator. Importantly, the procedure does not require any modification of the underlying synthetic control estimator: any synthetic-control-type (SC-type) method whose counterfactual is a weighted average of donor outcomes can be used, for example those proposed by \citet{Aba2019}, \citet{Amj2018}, \citet{ben2020}, \citet{Ben2019}, \citet{Dou2016}, \citet{Fer2019}, \citet{Kel2020} and \citet{Xu2017}. Beyond the assumptions of the chosen estimator, iSCM requires \textit{a priori} knowledge of which units are potentially affected by the intervention.

Our main result is stated in two parts: an exact finite-sample representation of the estimation error, which holds without any condition on the quality of the underlying synthetic control fits and shows that iSCM removes the contamination originating from the potentially affected units exactly, and a consistency statement obtained when the approximation errors of those fits vanish asymptotically. We complement the main result with a representation of iSCM as a weighted average of pure-control outcomes only, with implied weights that combine the $m$ unrestricted synthetic control fits. This representation clarifies what iSCM adds relative to the \textit{restricted} SCM that simply drops the potentially affected units: the implied iSCM weights inherit the pre-intervention information contained in the unrestricted fits, which the restricted optimizer, whose criterion changes with the donor pool, has no reason to recover. The advantage of iSCM over the restricted SCM is therefore governed by a trade-off between the information contained in the unrestricted fits and the amplification of their approximation errors in the correction step, and we provide a transparent comparison of the respective biases. When the restricted fit is comparable to the unrestricted one, this comparison simply indicates that the restricted SCM can be used and no correction is needed; our procedure identifies this case in the data.

The reliability of the correction depends on the conditioning of the linear system. The system becomes unstable when potentially affected units assign large and mutually reinforcing weights to one another, in which case the inversion step can amplify the approximation errors of the underlying fits; in such cases a \textit{simplified} version of the correction, which estimates the synthetic controls of the potentially affected units on the pure controls only and thereby avoids the inversion step, provides a useful alternative whenever those restricted fits remain adequate. The amplification concern is more severe when many units are classified as potentially affected. For this reason, we recommend imposing economically motivated restrictions on the set of potentially affected units, in line with the spillover literature, which typically allows interactions only within pre-specified groups (\citealp{Cer2017}, \citealp{For2018}, \citealp{Hub2019}, and \citealp{Vaz2017}). We also provide a data-driven procedure to guide the choice among the restricted SCM and the full and simplified versions of iSCM, together with practical guidance on the selection of the set of potentially affected units.

Our paper is related to the small literature extending SCM to settings with spillovers. \citet{grossi2024direct} estimate the effect on the treated unit by SCM with a restricted donor pool and recover spillover effects by comparing affected units with the restricted pool; their approach works well when the restricted pool produces a good synthetic control, but may be unreliable when the affected units are essential for the pre-intervention fit, as in the German reunification example. \citet{Cao2019}, like us, keep the potentially affected units in the analysis and estimate the direct and spillover effects jointly; their approach, however, imposes structure on the spillover mechanism and develops inference tailored to a specific estimator, whereas iSCM leaves the effects on the potentially affected units fully unrestricted, at the level of the individual unit and period, and operates on top of any SC-type estimator without changing its optimization problem. Both approaches require \textit{a priori} knowledge of the set of potentially affected units. In addition, iSCM covers applications in which other treated units must be included in the donor pool. Although related to the spillover literature, our results equally apply when there are multiple treated units and no spillovers. Our setting can also be viewed as a scenario with multiple treated units (\citealp{Aba2019}, \citealp{Ben2019}, and \citealp{Kel2020}). These methods modify the SCM objective, for example through penalization, or combine SCM with matching, but they are not designed for situations in which the ideal donor pool includes units potentially affected by the intervention. iSCM addresses precisely this case, without modifying the underlying estimator, and can in fact be combined with any of these methods.

The remainder of the paper is organized as follows. Section~\ref{sec:method} introduces the set-up, the assumptions and the main result, and summarizes the implementation of iSCM. Section~\ref{sec:purecontrol} derives the pure-control representation of iSCM and discusses its relation to restricted SCM. Section~\ref{sec:m2} studies the leading case of a single potentially affected unit, introduces the simplified iSCM and compares the biases of the unrestricted SCM, the full and simplified iSCM and the restricted SCM. Section~\ref{sec:guidance} provides practical guidance, including a data-driven procedure for choosing among the restricted SCM and the full and simplified iSCM and for selecting the set of potentially affected units. Section~\ref{sec:inference} discusses placebo-based inference. Section~\ref{sec:application} presents the empirical application to German reunification, and Section~\ref{sec:conclusion} concludes.

\section{The inclusive synthetic control method}\label{sec:method}
\setcounter{equation}{0}
\setcounter{assumption}{0}
\setcounter{lemma}{0}
\setcounter{theorem}{0}
\setcounter{remark}{0}

\subsection{Set-up and notation}\label{sec:setup}

Suppose we observe units $j=1,\ldots,J$ over periods $t=1,\ldots,T$, and that an intervention occurs at time $T_0<T$. For each unit $j$ we observe the outcome of interest $Y_{jt}$ in every period $t$, together with a $(k \times 1)$ vector of predictors $X_{j}$, which may include pre-intervention values of the outcome. Unit $1$, the \textit{main treated} unit, is directly exposed to the intervention. Among the remaining $J-1$ units, units $2$ to $m$, with $2\le m<J$, are \textit{potentially affected} by the intervention, directly or indirectly: they are either other treated units that the researcher would like to keep in the donor pool, or control units that may be exposed to spillover effects originating from the intervention. We collect their indices in $M\equiv\{2,\ldots,m\}$, so that the analysis involves $m$ unknown effects, namely the effect on the main treated unit and the $m-1$ effects on the units in $M$. Units $m+1$ to $J$ are \textit{pure controls}, collected in $C\equiv\{m+1,\ldots,J\}$, and are assumed to be entirely unaffected by the intervention. We let $Y_{Ct}\equiv (Y_{jt})_{j\in C}$ collect the pure-control outcomes at time $t$.

We adopt the potential outcome framework. Let $Y_{1t}^{I}$ denote the potential outcome of the main treated unit under the intervention, and let $Y_{jt}^{S}$, $j\in M$, denote the potential outcome of a potentially affected unit, which represents either the outcome of a unit exposed to spillover effects or the outcome under treatment when unit $j$ is itself treated. Finally, let $Y_{jt}^{N}$, $j=1,\ldots,J$, denote the potential outcome in the absence of the intervention. We assume that potential outcomes are indexed by a unit's own exposure to the intervention only, so that the only admissible exposures are those of the units in $\{1\}\cup M$; this is a partial-interference, or exposure-mapping, restriction of the type commonly imposed in the spillover literature. In the assumption below, part (a) is a no-anticipation condition, and part (b) is the key restriction that rules out effects on units outside $\{1\}\cup M$; the units in $M$ are chosen \textit{a priori}.

\begin{assumption}\label{ass:sutva}
(a) $Y_{jt}=Y_{jt}^{N}$ for all $j=1,\ldots,J$ and all $t\le T_0$;
(b) $Y_{jt}=Y_{jt}^{N}$ for all $j\in C$ and all $t> T_0$;
(c) $Y_{1t}=Y_{1t}^{I}$ for all $t> T_0$;
(d) $Y_{jt}=Y_{jt}^{S}$ for all $j\in M$ and all $t> T_0$.
\end{assumption}

Assumption~\ref{ass:sutva} may be restrictive in applications where it is unclear which units could be affected by spillovers; it is, however, standard when the potentially affected units are other treated units. Section~\ref{sec:choosingM} discusses how to choose the set $M$ in practice.

We are interested in the causal effect of the intervention on the main treated unit and in the effects on the potentially affected units,
\begin{equation}\label{eq:estimands}
\theta_{1t}=Y_{1t}^{I}-Y_{1t}^{N}
\qquad\text{and}\qquad
\gamma_{jt}=Y_{jt}^{S}-Y_{jt}^{N}, \quad j\in M,
\end{equation}
for $t>T_0$. We impose no restrictions on $\theta_{1t}$ or $\gamma_{jt}$: the effects may differ arbitrarily across units and over time, and no additive or uniform spillover structure is assumed. Recovering these effects requires estimates of the unobserved counterfactuals $Y_{jt}^{N}$, $j=1,\ldots,m$, in the post-intervention period.

We consider synthetic-control-type (SC-type) estimators, that is, estimators that approximate the counterfactual of a given unit by a weighted average of donor outcomes, with weights estimated using pre-intervention data only. This class includes the original SCM of \citet{Aba2010} as well as, among others, the estimators proposed by \citet{Aba2019}, \citet{Amj2018}, \citet{ben2020}, \citet{Ben2019}, \citet{Dou2016}, \citet{Fer2019}, \citet{Kel2020} and \citet{Xu2017}. For the main treated unit, the donor pool consists of all remaining units, including the potentially affected ones, and the estimated counterfactual is
\begin{equation}\label{eq:scmain}
\widehat{Y}_{1t}^{N}=\sum_{j=2}^{J}\widehat{w}_{j}Y_{jt},
\end{equation}
where $\widehat{W}=(\widehat{w}_{2},\ldots,\widehat{w}_{J})'$ is the $((J-1)\times 1)$ vector of weights produced by the chosen SC-type estimator. As a leading example, the original SCM selects $\widehat{W}$ by minimizing the predictor imbalance $\|X_{1}-X_{0}W\|_{V}$ over nonnegative weights that sum to one, where $X_{0}\equiv(X_{2},\ldots,X_{J})$ collects the donor predictors and $V$ is a positive semidefinite weighting matrix chosen by the data-driven procedure of \citet{Aba2010} or by the sample-splitting method of \citet{Aba2015}. Similarly, for each potentially affected unit $i\in M$, the donor pool consists of all units other than $i$, including the main treated unit, and
\begin{equation}\label{eq:scaffected}
\widehat{Y}_{it}^{N}=\sum_{j\neq i}\widehat{l}^{\,i}_{j}Y_{jt},
\end{equation}
where $\widehat{L}^{i}$ collects the corresponding weights. The associated effect estimators are
\begin{equation}\label{eq:gaps}
\widehat{\theta}_{1t}\equiv Y_{1t}-\widehat{Y}_{1t}^{N}
\qquad\text{and}\qquad
\widehat{\gamma}_{it}\equiv Y_{it}-\widehat{Y}_{it}^{N},\quad i\in M, \quad t>T_0.
\end{equation}

\subsection{Two sources of estimation error}\label{sec:decomposition}

Under Assumption~\ref{ass:sutva}, for every $t>T_0$ the estimated counterfactual of the main treated unit satisfies
\begin{eqnarray*}
\widehat{Y}_{1t}^{N}&=&\sum_{j=m+1}^{J}\widehat{w}_{j}Y_{jt}+\sum_{j=2}^{m}\widehat{w}_{j}Y_{jt}
=\sum_{j=m+1}^{J}\widehat{w}_{j}Y_{jt}^{N}+\sum_{j=2}^{m}\widehat{w}_{j}\big(Y_{jt}^{N}+\gamma_{jt}\big)\\
&=&\sum_{j=2}^{J}\widehat{w}_{j}Y_{jt}^{N}+\sum_{j=2}^{m}\widehat{w}_{j}\gamma_{jt}.
\end{eqnarray*}
The estimation error of $\widehat{Y}_{1t}^{N}$ can therefore be decomposed into two parts:
\begin{equation}\label{eq:dec1}
\widehat{Y}_{1t}^{N}-Y_{1t}^{N}
=\underbrace{\bigg(\sum_{j=2}^{J}\widehat{w}_{j}Y_{jt}^{N}-Y_{1t}^{N}\bigg)}_{B^{1}_{sc}}
+\underbrace{\bigg(\sum_{j=2}^{m}\widehat{w}_{j}\gamma_{jt}\bigg)}_{B^{1}_{te}}.
\end{equation}
Similarly, for a generic potentially affected unit $i\in M$, Assumption~\ref{ass:sutva} implies
\begin{eqnarray*}
\widehat{Y}_{it}^{N}&=&\sum_{j=m+1}^{J}\widehat{l}^{\,i}_{j}Y_{jt}+\sum_{j\in M\setminus\{i\}}\widehat{l}^{\,i}_{j}Y_{jt}+\widehat{l}^{\,i}_{1}Y_{1t}\\
&=&\sum_{j\neq i}\widehat{l}^{\,i}_{j}Y_{jt}^{N}+\sum_{j\in M\setminus\{i\}}\widehat{l}^{\,i}_{j}\gamma_{jt}+\widehat{l}^{\,i}_{1}\theta_{1t},
\end{eqnarray*}
so that
\begin{equation}\label{eq:dec2}
\widehat{Y}_{it}^{N}-Y_{it}^{N}
=\underbrace{\bigg(\sum_{j\neq i}\widehat{l}^{\,i}_{j}Y_{jt}^{N}-Y_{it}^{N}\bigg)}_{B^{i}_{sc}}
+\underbrace{\bigg(\sum_{j\in M\setminus\{i\}}\widehat{l}^{\,i}_{j}\gamma_{jt}+\widehat{l}^{\,i}_{1}\theta_{1t}\bigg)}_{B^{i}_{te}},
\quad i\in M.
\end{equation}

The decompositions in \eqref{eq:dec1} and \eqref{eq:dec2} isolate two conceptually distinct sources of estimation error which, with a slight abuse of terminology, we refer to as biases throughout. The first, $B^{j}_{sc}$, $j=1,\ldots,m$, is the approximation error of the chosen SC-type estimator, i.e.\ how well the weighted average of untreated potential outcomes reproduces the counterfactual of interest. This is the usual source of bias of any SC-type estimator. The second, $B^{j}_{te}$, $j=1,\ldots,m$, arises solely because units that are potentially affected by the intervention enter the donor pool: it depends only on the weights assigned to those units and on their treatment or spillover effects, not on the quality of the chosen estimator. The iSCM is designed to remove this second source of bias.

\subsection{The iSCM estimator}\label{sec:mainresult}

Throughout, we study the estimators of Section~\ref{sec:setup} under an asymptotic sequence in which the number of pre-intervention periods $T_{0}$ diverges while the number of units $J$, and hence $m$, is fixed. Formally, all random objects are defined on a common probability space, and the analysis proceeds along a triangular array indexed by $T_{0}$: for each $T_{0}$, the estimated weights are functions of the pre-intervention sample, and the post-intervention periods are indexed by a fixed horizon $h=t-T_{0}\ge 1$. The effects $\theta_{1,T_{0}+h}$ and $\gamma_{j,T_{0}+h}$ may be treated as fixed or random; all $o_{p}(1)$ statements are with respect to this sequence and are pointwise in $h$, and since the number of post-intervention periods is finite, pointwise and uniform statements coincide. We deliberately state our conditions at the level of the estimated weights, as high-level requirements on the chosen SC-type estimator. This keeps the analysis applicable to the whole class of estimators above and separates the contamination problem, which is our focus, from the approximation properties of any specific estimator. Primitive conditions for the original SCM are studied by \citet{Zha2022}. We stress, however, that under a linear factor model with non-degenerate idiosyncratic shocks, conditions of this form hold at best approximately at any fixed horizon, because the weighted idiosyncratic errors do not vanish, even when the estimated weights are consistent for the pseudo-true weights that match the factor loadings (\citealp{Aba2010}; \citealp{Fer2019}). For this reason, our main result is stated in two parts: an exact finite-sample representation of the estimation error, which holds without any restriction on the approximation errors of the underlying fits and is the operative result under such data-generating processes, and a consistency statement obtained when those errors are asymptotically negligible.

We first assume that the chosen SC-type estimator approximates the counterfactual of the main treated unit well, in the sense that the approximation error $B^{1}_{sc}$ is asymptotically negligible.

\begin{assumption}\label{ass:fitmain}
For every fixed $h\ge 1$, $\sum_{j=2}^{J}\widehat{w}_{j}Y_{j,T_{0}+h}^{N}=Y_{1,T_{0}+h}^{N}+o_{p}(1)$ as $T_{0}\rightarrow\infty$.
\end{assumption}

Assumption~\ref{ass:fitmain} is stated in terms of the estimated weights, as a high-level condition and to simplify the exposition; alternatively, one can assume that $\widehat{W}$ converges to a vector of weights $W^{*}$ such that $\sum_{j=2}^{J}w^{*}_{j}Y_{j,T_{0}+h}^{N}=Y_{1,T_{0}+h}^{N}+o_{p}(1)$. Under a factor-model justification, Assumption~\ref{ass:fitmain} is commonly motivated by an (approximately) perfect pre-intervention fit over a long pre-intervention period; a good pre-intervention fit is, however, neither necessary nor sufficient for reproducing the post-intervention counterfactual without further structure.\footnote{See the discussions in \citet{Aba2020}, \citet{Fer2019} and \citet{Pow2022} on the consequences of a lack of perfect fit.} The assumption can be weakened by employing an SC-type estimator that is robust to violations of the perfect fit condition, such as those of \citet{ben2020} or \citet{Fer2019}, and can be stated more primitively for specific estimators, e.g.\ using the results of \citet{Zha2022} for the original SCM.

\begin{lemma}\label{lem:main}
Under Assumptions \ref{ass:sutva} and \ref{ass:fitmain}, for every fixed $h\ge 1$,
\[
\widehat{\theta}_{1,T_{0}+h}=\theta_{1,T_{0}+h}-\sum_{j=2}^{m}\widehat{w}_{j}\gamma_{j,T_{0}+h}+o_{p}(1)
\qquad\text{as } T_{0}\rightarrow\infty.
\]
\end{lemma}

\noindent\textbf{Proof:} Immediate from \eqref{eq:dec1}, Assumption~\ref{ass:fitmain} and $\widehat{\theta}_{1t}=\theta_{1t}-(\widehat{Y}_{1t}^{N}-Y_{1t}^{N})$.\hfill$\square$

\begin{remark}\label{rem:smallweights}
\textnormal{For each $j\in M$, if either $\gamma_{jt}$ or $\widehat{w}_{j}$ is zero, unit $j$ induces no contamination bias in $\widehat{\theta}_{1t}$. Units receiving a small weight would need an extremely large effect to induce a non-negligible bias, so such units can be treated as pure controls relatively safely when the object of interest is $\theta_{1t}$. When using SC-type estimators that allow for negative weights, the magnitude of the weights assigned to potentially affected units should be checked explicitly.}
\end{remark}

\begin{remark}\label{rem:biascorrected}
\textnormal{For estimators that incorporate a bias correction, such as the bias-corrected estimators discussed in \citet{Aba2020} (see equation (16) therein) and \citet{Aba2019}, some care is required. The correction regressions are typically fitted on post-intervention donor outcomes, so if units in $M$ enter as donors, their post-intervention outcomes embed the effects $\gamma_{jt}$ and the fitted correction is itself contaminated. Two modifications restore the logic of iSCM: (a) fit the bias-correction regressions on the pure controls only, at the cost of a smaller estimation sample; or (b) proceed iteratively, first estimating the effects by iSCM without bias correction and then subtracting the estimated effects from the outcomes of the units in $M$ before fitting the correction. We recommend option (a) for its simplicity.}
\end{remark}

Next, we require the analogous condition for the potentially affected units.

\begin{assumption}\label{ass:fitaffected}
For every $i\in M$ and every fixed $h\ge 1$, $\sum_{j\neq i}\widehat{l}^{\,i}_{j}Y_{j,T_{0}+h}^{N}=Y_{i,T_{0}+h}^{N}+o_{p}(1)$ as $T_{0}\rightarrow\infty$.
\end{assumption}

Assumption~\ref{ass:fitaffected} requires the chosen estimator to deliver a credible counterfactual for every potentially affected unit, whereas applied researchers typically scrutinize the fit of the main treated unit only. The plausibility of Assumption~\ref{ass:fitaffected} can, however, be assessed empirically: the fit of every unit in $M$ is observable in the pre-intervention period and should be examined in the same way as the fit of the main treated unit (see Section~\ref{sec:guidance}). When a satisfactory fit cannot be achieved for some unit in $M$, the consequences of removing it should be acknowledged explicitly: excluding it from all donor pools removes its contamination at the cost of a (partially) restricted analysis, whereas reclassifying it as a pure control is harmless only when it receives negligible weight (Remark~\ref{rem:smallweights}) or when its effect is plausibly small (see Section~\ref{sec:choosingM}).

\begin{lemma}\label{lem:affected}
Under Assumptions \ref{ass:sutva} and \ref{ass:fitaffected}, for every $i\in M$ and every fixed $h\ge 1$,
\[
\widehat{\gamma}_{i,T_{0}+h}=\gamma_{i,T_{0}+h}-\sum_{j\in M\setminus\{i\}}\widehat{l}^{\,i}_{j}\gamma_{j,T_{0}+h}-\widehat{l}^{\,i}_{1}\theta_{1,T_{0}+h}+o_{p}(1)
\qquad\text{as } T_{0}\rightarrow\infty.
\]
\end{lemma}

\noindent\textbf{Proof:} Immediate from \eqref{eq:dec2}, Assumption~\ref{ass:fitaffected} and $\widehat{\gamma}_{it}=\gamma_{it}-(\widehat{Y}_{it}^{N}-Y_{it}^{N})$.\hfill$\square$

The decompositions \eqref{eq:dec1} and \eqref{eq:dec2} show that each contaminated estimate equals a known linear combination of the $m$ unknown effects, up to the corresponding approximation error. Treating the effects as unknowns and omitting the approximation errors for readability, we obtain the system
\begin{eqnarray*}
\widehat{\theta}_{1t}&=&\theta_{1t}-\widehat{w}_{2}\gamma_{2t}-\widehat{w}_{3}\gamma_{3t}-\ldots-\widehat{w}_{m}\gamma_{mt},\\
\widehat{\gamma}_{2t}&=&-\widehat{l}^{\,2}_{1}\theta_{1t}+\gamma_{2t}-\widehat{l}^{\,2}_{3}\gamma_{3t}-\ldots-\widehat{l}^{\,2}_{m}\gamma_{mt},\\
&\vdots&\\
\widehat{\gamma}_{mt}&=&-\widehat{l}^{\,m}_{1}\theta_{1t}-\widehat{l}^{\,m}_{2}\gamma_{2t}-\widehat{l}^{\,m}_{3}\gamma_{3t}-\ldots+\gamma_{mt}.
\end{eqnarray*}
In matrix form, let $\vartheta_{t}$ denote the $(m\times 1)$ vector of effects of interest, $\widehat{\beta}_{t}$ the $(m\times 1)$ vector of contaminated estimates, and $\widehat{\Omega}$ the $(m\times m)$ matrix of estimated weights with ones on the main diagonal:
\begin{equation}\label{eq:system}
\widehat{\beta}_{t}=
\begin{pmatrix}
\widehat{\theta}_{1t}\\
\widehat{\gamma}_{2t}\\
\vdots\\
\widehat{\gamma}_{mt}
\end{pmatrix},
\qquad
\widehat{\Omega}=
\begin{pmatrix}
1 & -\widehat{w}_{2} & \dots & -\widehat{w}_{m}\\
-\widehat{l}^{\,2}_{1} & 1 & \dots & -\widehat{l}^{\,2}_{m}\\
\vdots & \vdots & \ddots & \vdots\\
-\widehat{l}^{\,m}_{1} & -\widehat{l}^{\,m}_{2} & \dots & 1
\end{pmatrix},
\qquad
\vartheta_{t}=
\begin{pmatrix}
\theta_{1t}\\
\gamma_{2t}\\
\vdots\\
\gamma_{mt}
\end{pmatrix}.
\end{equation}
Reinstating the approximation errors, stacking the decompositions \eqref{eq:dec1} and \eqref{eq:dec2} yields the exact relation
\begin{equation}\label{eq:exactrelation}
\widehat{\beta}_{t}=\widehat{\Omega}\,\vartheta_{t}-B_{t},
\qquad
B_{t}\equiv\big(B^{1}_{sc},\ldots,B^{m}_{sc}\big)',
\qquad t>T_{0},
\end{equation}
which holds under Assumption~\ref{ass:sutva} alone; the system above is \eqref{eq:exactrelation} with $B_{t}$ omitted.

Solving the system requires $\widehat{\Omega}$ to be invertible and, because $\widehat{\beta}_{t}$ and $\widehat{\Omega}$ are estimated, well conditioned. The finite-sample statement below requires only nonsingularity of the realized matrix; for the asymptotic statement we impose a minimum singular value condition.

\begin{assumption}\label{ass:invertibility}
$P\big(\sigma_{\min}(\widehat{\Omega})\ge c\big)\rightarrow 1$ as $T_{0}\rightarrow\infty$ for some constant $c>0$, where $\sigma_{\min}(\widehat{\Omega})$ denotes the smallest singular value of $\widehat{\Omega}$.
\end{assumption}

Assumption~\ref{ass:invertibility} requires the system to remain well conditioned asymptotically, so that the inversion step does not amplify estimation errors unduly: on the event $\{\sigma_{\min}(\widehat{\Omega})\ge c\}$, $\|\widehat{\Omega}^{-1}\|\le 1/c$ in spectral norm. The realized value of $\sigma_{\min}(\widehat{\Omega})$ is computable from the estimated weights and should be reported as a diagnostic; it cannot, of course, verify an asymptotic condition, but it bounds the amplification in the finite-sample statement below. Lemma~A.1 in Appendix~A shows that, whenever the underlying SC-type estimator imposes nonnegative weights that sum to one and every fit places strictly positive total weight on pure controls, $\widehat{\Omega}$ is strictly diagonally dominant and hence nonsingular in every realization; failures arise only in knife-edge cases (see Section~\ref{sec:m2} and Appendix~A for examples). Even when $\widehat{\Omega}$ is invertible, it may be nearly singular if the potentially affected units assign large weights to one another; in that case the correction can amplify the approximation errors $B^{j}_{sc}$, an issue we examine in detail in Sections \ref{sec:m2} and \ref{sec:guidance}.

\begin{theorem}\label{thm:main}
Let Assumption \ref{ass:sutva} hold, and define $\widehat{\vartheta}^{\,iSCM}_{t}\equiv\widehat{\Omega}^{-1}\widehat{\beta}_{t}$ whenever $\widehat{\Omega}$ is nonsingular.
(a) For every $t>T_{0}$, on the event that $\widehat{\Omega}$ is nonsingular, $\widehat{\vartheta}^{\,iSCM}_{t}-\vartheta_{t}=-\widehat{\Omega}^{-1}B_{t}$, and hence $\|\widehat{\vartheta}^{\,iSCM}_{t}-\vartheta_{t}\|\le\|B_{t}\|/\sigma_{\min}(\widehat{\Omega})$, where $\|\cdot\|$ denotes the Euclidean norm for vectors and the spectral norm for matrices.
(b) If, in addition, Assumptions \ref{ass:fitmain}, \ref{ass:fitaffected} and \ref{ass:invertibility} hold, then for every fixed $h\ge 1$, $\widehat{\vartheta}^{\,iSCM}_{T_{0}+h}=\vartheta_{T_{0}+h}+o_{p}(1)$ as $T_{0}\rightarrow\infty$.
\end{theorem}

\noindent\textbf{Proof:} Premultiplying \eqref{eq:exactrelation} by $\widehat{\Omega}^{-1}$, on the event that $\widehat{\Omega}$ is nonsingular, gives part (a); the bound follows because $\|\widehat{\Omega}^{-1}\|=1/\sigma_{\min}(\widehat{\Omega})$ in spectral norm. For part (b), Assumptions \ref{ass:fitmain} and \ref{ass:fitaffected} state precisely that $B_{T_{0}+h}=o_{p}(1)$, while Assumption~\ref{ass:invertibility} implies that $\|\widehat{\Omega}^{-1}\|\le 1/c$ with probability approaching one, so that $\widehat{\Omega}^{-1}B_{T_{0}+h}=o_{p}(1)$.\hfill$\square$

We write $\widehat{\vartheta}^{\,iSCM}_{t}=(\widehat{\theta}^{\,iSCM}_{1t},\widehat{\gamma}^{\,iSCM}_{2t},\ldots,\widehat{\gamma}^{\,iSCM}_{mt})'$ for the components of the iSCM estimator. Part (a) of Theorem~\ref{thm:main} holds in finite samples and without any condition on the quality of the underlying fits: iSCM removes the contamination biases $B^{j}_{te}$ exactly, and the error that remains is the vector of approximation errors $B_{t}$, which includes any idiosyncratic deviations from the fit conditions, propagated through $\widehat{\Omega}^{-1}$ with an amplification bounded by $1/\sigma_{\min}(\widehat{\Omega})$, a quantity computable from the estimated weights that should be reported in applications. Part (a) therefore describes precisely how the estimator behaves under imperfect fit and, as discussed at the beginning of this subsection, it is the operative result under a linear factor model with non-degenerate idiosyncratic shocks; Section~\ref{sec:biascomparison} specializes it to $m=2$, where the amplification takes a transparent form. Part (b) adds that the estimator is consistent for the effects of interest whenever the approximation errors vanish asymptotically.

The iSCM estimator is obtained by solving an $m$-dimensional linear system, an operation that requires only basic linear algebra in addition to the chosen SC-type estimator. When $m=2$, the solution has a simple closed form, which we exploit in Section~\ref{sec:m2}; in general, we recommend solving the system directly rather than computing individual entries (e.g., by Cramer's rule).\footnote{By Cramer's rule, the $k$-th component of $\widehat{\vartheta}^{\,iSCM}_{t}$ equals $\det(\widehat{\Omega}_{(k)})/\det(\widehat{\Omega})$, where $\widehat{\Omega}_{(k)}$ denotes $\widehat{\Omega}$ with its $k$-th column replaced by $\widehat{\beta}_{t}$; for $m=2$ this delivers the closed form in \eqref{eq:tiscm}. While useful for interpretation, computing individual entries in this way is numerically inferior to solving the system directly.}

\subsection{Implementation}\label{sec:implementation}

For convenience, we summarize the steps required to compute the iSCM estimator; all objects are defined in Sections \ref{sec:setup}--\ref{sec:mainresult}.

\setcounter{bean}{0}
\begin{center}
\begin{list}
{\textsc{Step} \arabic{bean}.}{\usecounter{bean}}
\item Using pre-intervention data only, estimate the SC-type weights $\widehat{W}$ for the main treated unit with all remaining units in the donor pool; compute $\widehat{Y}^{N}_{1t}=\sum_{j=2}^{J}\widehat{w}_{j}Y_{jt}$ and $\widehat{\theta}_{1t}=Y_{1t}-\widehat{Y}^{N}_{1t}$ for $t>T_{0}$.
\item For each $i\in M$, estimate the SC-type weights $\widehat{L}^{i}$ for unit $i$ with all units other than $i$ in the donor pool; compute $\widehat{Y}^{N}_{it}=\sum_{j\neq i}\widehat{l}^{\,i}_{j}Y_{jt}$ and $\widehat{\gamma}_{it}=Y_{it}-\widehat{Y}^{N}_{it}$ for $t>T_{0}$.
\item Form $\widehat{\beta}_{t}=(\widehat{\theta}_{1t},\widehat{\gamma}_{2t},\ldots,\widehat{\gamma}_{mt})'$ and the matrix $\widehat{\Omega}$ in \eqref{eq:system}, and check that $\widehat{\Omega}$ is well conditioned, by computing $\sigma_{\min}(\widehat{\Omega})$ or checking strict diagonal dominance (see Lemma~A.1 in Appendix~A and Remark~\ref{rem:diagdominance}).
\item Compute $\widehat{\vartheta}^{\,iSCM}_{t}=\widehat{\Omega}^{-1}\widehat{\beta}_{t}$ for $t>T_{0}$.
\end{list}
\end{center}

These steps deliver the full iSCM estimator. The choice among the restricted SCM and the full and simplified iSCM, together with the diagnostics on which it is based, is discussed in Section~\ref{sec:guidance}. An R package, \texttt{iscm}, implementing Steps 1--4 above, the diagnostics of Section~\ref{sec:guidance} and the placebo procedure of Section~\ref{sec:inference}, together with a replication script for the empirical application of Section~\ref{sec:application}, is available from the authors.

\section{A pure-control representation}\label{sec:purecontrol}
\setcounter{equation}{0}
\setcounter{lemma}{0}
\setcounter{remark}{0}

The iSCM correction removes the post-intervention contamination originating from the potentially affected units. In this section we make this explicit by rewriting the iSCM estimate for the main treated unit as a linear combination of pure-control outcomes only. The representation shows precisely what is gained by keeping the potentially affected units in the donor pool: the implied weights inherit the pre-intervention information of the $m$ unrestricted fits, which the restricted SCM, whose optimization problem changes with the donor pool, has no reason to recover. It thereby clarifies the relation between iSCM and the restricted SCM that excludes the potentially affected units from the donor pool, and it underlies the bias comparison of Section~\ref{sec:m2} and the decision rule of Section~\ref{sec:guidance}.

For each $i\in\{1,\ldots,m\}$, let $\widehat{p}_{i}\in\mathbb{R}^{|C|}$ denote the vector of weights that the $i$-th SC-type fit assigns to the pure-control units: for $i=1$ these are the components of $\widehat{W}$ corresponding to $C$, and for $i\in M$ they are the components of $\widehat{L}^{i}$ corresponding to $C$. Stack these vectors row-wise in the $(m\times |C|)$ matrix $\widehat{P}\equiv(\widehat{p}_{1},\ldots,\widehat{p}_{m})'$ and let $\widehat{s}\in\mathbb{R}^{m}$ collect the total pure-control weights $\widehat{s}_{i}\equiv\mathbf{1}_{C}'\widehat{p}_{i}$, where $\mathbf{1}_{m}$ and $\mathbf{1}_{C}$ denote vectors of ones of dimensions $m$ and $|C|$, respectively. Finally, let $e_{1}$ denote the first canonical basis vector in $\mathbb{R}^{m}$.

\begin{lemma}\label{lem:purecontrol}
Whenever $\widehat{\Omega}$ is nonsingular, for each $t>T_{0}$,
\begin{equation}\label{eq:purecontrolrep}
\widehat{\theta}^{\,iSCM}_{1t}=Y_{1t}-Y_{Ct}'\,\widehat{a},
\qquad
\widehat{a}\equiv\big(e_{1}'\widehat{\Omega}^{-1}\widehat{P}\big)'.
\end{equation}
If, in addition, the underlying SC-type estimator imposes adding-up constraints, so that the weights in each of the $m$ fits sum to one, then $\mathbf{1}_{C}'\widehat{a}=1$. If, moreover, (a) $\widehat{P}\ge 0$ elementwise and (b) $\widehat{\Omega}$ is a nonsingular M-matrix, so that $\widehat{\Omega}^{-1}\ge 0$ elementwise, then $\widehat{a}\ge 0$, and \eqref{eq:purecontrolrep} is a convex combination of pure-control outcomes.
\end{lemma}

Lemma~\ref{lem:purecontrol}, whose proof is in Appendix~A, formalizes a property that any procedure robust to arbitrary effects on the potentially affected units must possess: \textit{ex post}, only pure-control outcomes can enter the counterfactual, because the post-intervention outcomes of the units in $M$ are contaminated by unknown effects. The content of the lemma is the explicit construction of the implied weights $\widehat{a}$, which combine the $m$ unrestricted fits through $\widehat{\Omega}^{-1}$.

\begin{remark}\label{rem:diagdominance}
\textnormal{A convenient sufficient condition for $\widehat{\Omega}$ to be a nonsingular M-matrix is strict diagonal dominance by rows, $\sum_{j\neq i}|\widehat{\Omega}_{ij}|<1$ for all $i=1,\ldots,m$. When the SC-type estimator imposes nonnegativity and adding-up constraints, this corresponds to each fit placing strictly positive total weight on the pure controls, $\widehat{s}_{i}>0$; see Lemma~A.1 in Appendix~A. Both conditions can be checked directly from the estimated weights.}
\end{remark}

\begin{remark}\label{rem:polarcases}
\textnormal{Two polar cases are instructive. (a) When there is a single pure-control unit, $|C|=1$, the adding-up property forces $\widehat{a}=1$, and iSCM coincides with the restricted SCM, which assigns weight one to the single pure control. (b) When every fit for the potentially affected units places zero weight on the main treated unit and on the other units in $M$, the rows $2,\ldots,m$ of $\widehat{\Omega}$ reduce to unit vectors, and \eqref{eq:purecontrolrep} simplifies to $\widehat{a}=\widehat{p}_{1}+\sum_{j\in M}\widehat{w}_{j}\widehat{p}_{j}$. In this case iSCM is numerically equivalent to the intuitive plug-in approach that replaces each potentially affected unit in the synthetic control of the main treated unit with that unit's own synthetic control constructed from pure controls; Section~\ref{sec:m2} refers to the estimator that imposes this structure deliberately as the simplified iSCM. The general construction in Theorem~\ref{thm:main} extends this logic to the case in which the main treated unit and the other affected units are needed in the donor pools of the units in $M$.}
\end{remark}

\subsection{Relation to restricted SCM}\label{sec:relationrestricted}

Lemma~\ref{lem:purecontrol} raises a natural question: if iSCM ultimately aggregates pure-control outcomes only, why not simply apply the chosen SC-type estimator to the restricted donor pool $C$? The question is sharpened by the observation that, under the convexity conditions of Lemma~\ref{lem:purecontrol}, the implied weights $\widehat{a}$ are nonnegative and sum to one, and are therefore \textit{feasible} for the restricted problem. The answer lies in the selection criterion. SC-type estimators do not select weights by minimizing the post-intervention, or even the pre-intervention, outcome discrepancy directly: the original SCM, for instance, minimizes the predictor imbalance $\|X_{1}-X_{0}W\|_{V}$, with the weighting matrix $V$ itself re-estimated for each donor pool, and other SC-type estimators use penalized or augmented criteria. Changing the donor pool therefore changes the entire optimization problem, and nothing guarantees that the restricted optimizer comes close to $\widehat{a}$, even though $\widehat{a}$ lies in its feasible set. Our application illustrates the gap. For $m=2$, the pre-intervention gap associated with the implied weights equals $\alpha(\widehat{e}_{1t}+\widehat{w}_{2}\widehat{e}_{2t})$, where $\widehat{e}_{1t}$ and $\widehat{e}_{2t}$ denote the pre-intervention residuals of the two unrestricted fits and $\alpha$ is the amplification factor defined in Section~\ref{sec:biascomparison}. By the triangle inequality, the pre-intervention root mean squared prediction error (RMSPE) of the implied weights therefore satisfies the upper bound
\begin{equation}\label{eq:rmspebound}
\mathrm{RMSPE}\big(\widehat{a}\big)\le\alpha\big(\mathrm{RMSPE}_{1}+\widehat{w}_{2}\,\mathrm{RMSPE}_{2}\big),
\end{equation}
where $\mathrm{RMSPE}_{1}$ and $\mathrm{RMSPE}_{2}$ denote the pre-intervention RMSPEs of the two unrestricted fits. In our application the bound evaluates to approximately $1.16\times(119.07+0.42\times 194.67)\approx 233$, and the exact pre-intervention RMSPE of the implied weights is 138.54; the ingredients are reported in Section~\ref{sec:application}, where the calculation is restated. The restricted optimizer instead attains an RMSPE of 260.74: the weights it selects fit the outcome markedly worse than the implied weights, which lie in its own feasible set, because it targets a different criterion. Note that, for $t\le T_{0}$, figures in applied work, including ours, conventionally plot the gap of the unrestricted fit rather than that of the implied weights, since the correction is defined only for $t>T_{0}$; the pre-intervention fit of the implied weights in \eqref{eq:rmspebound} is a distinct object, relevant for the comparison with the restricted SCM. The two procedures coincide only in special cases, such as that of Remark~\ref{rem:polarcases}(a), and conversely the restricted optimizer cannot exploit the information contained in the pre-intervention trajectories of the units in $M$.

The substantive difference is therefore in the approximation errors that each construction inherits. The bias of the restricted SCM is governed by the out-of-sample approximation error of the weights its optimizer selects on the restricted pool, which can be large when the potentially affected units are the closest to the main treated unit, geographically or economically, and the restricted criterion is forced to rely on more distant donors. The bias of iSCM is instead governed by the approximation errors of the $m$ unrestricted fits, propagated through $\widehat{\Omega}^{-1}$ as in Theorem~\ref{thm:main}(a); each of these errors may be small precisely because the donor pools are unrestricted, but their amplification grows with the cross-weights among the units in $M$. Neither construction dominates: when the system is well conditioned and the affected units carry information that the restricted optimizer cannot recover, iSCM is preferable; when the cross-weights are large, the amplification can offset the informational gain. Section~\ref{sec:m2} makes this comparison precise in the leading case $m=2$, and Section~\ref{sec:guidance} translates it into a data-driven decision rule. Appendix~C complements this discussion with a geometric illustration: excluding the potentially affected units can push the main treated unit outside the convex hull of the remaining donors, or force the restricted fit to rely on units far from the main treated unit, increasing interpolation bias (\citealp{Aba2020}).

In sum, the pure-control representation shows in what precise sense iSCM combines the robustness of a restricted aggregation, which uses uncontaminated outcomes only, with the pre-intervention information of the unrestricted fits.

\section{The case of a single potentially affected unit}\label{sec:m2}
\setcounter{equation}{0}
\setcounter{proposition}{0}
\setcounter{corollary}{0}
\setcounter{remark}{0}

To build intuition, we now consider the special case in which only one unit, in addition to the main treated unit, is potentially affected by the intervention, i.e.\ $m=2$. To lighten the notation we write $\theta_{t}$ and $\gamma_{t}$ for $\theta_{1t}$ and $\gamma_{2t}$, and $\widehat{l}_{1}$ for $\widehat{l}^{\,2}_{1}$, the weight that unit 2's fit assigns to the main treated unit; as before, $\widehat{w}_{2}$ denotes the weight that the main treated unit's fit assigns to unit 2.

\subsection{Closed-form solution}

With $m=2$ and the effects again treated as unknowns, the system of Section~\ref{sec:mainresult} reduces to
\begin{eqnarray*}
\widehat{\theta}_{t}&=&\theta_{t}-\widehat{w}_{2}\gamma_{t},\\
\widehat{\gamma}_{t}&=&-\widehat{l}_{1}\theta_{t}+\gamma_{t},
\end{eqnarray*}
with
\begin{align*}
\widehat{\beta}_{t}=
\begin{pmatrix}
\widehat{\theta}_{t}\\
\widehat{\gamma}_{t}
\end{pmatrix},
\qquad
\widehat{\Omega}=
\begin{pmatrix}
1 & -\widehat{w}_{2}\\
-\widehat{l}_{1} & 1
\end{pmatrix},
\qquad
\vartheta_{t}=
\begin{pmatrix}
\theta_{t}\\
\gamma_{t}
\end{pmatrix}.
\end{align*}
Since $\det(\widehat{\Omega})=1-\widehat{w}_{2}\widehat{l}_{1}$, the system has the closed-form solution
\begin{equation}\label{eq:tiscm}
\widehat{\theta}^{\,iSCM}_{t}=\frac{\widehat{\theta}_{t}+\widehat{w}_{2}\widehat{\gamma}_{t}}{1-\widehat{w}_{2}\widehat{l}_{1}},
\qquad
\widehat{\gamma}^{\,iSCM}_{t}=\frac{\widehat{\gamma}_{t}+\widehat{l}_{1}\widehat{\theta}_{t}}{1-\widehat{w}_{2}\widehat{l}_{1}}.
\end{equation}
Under nonnegativity and adding-up constraints, the determinant vanishes only when $\widehat{w}_{2}=\widehat{l}_{1}=1$, i.e.\ when the main treated unit assigns weight one to the potentially affected unit and the potentially affected unit assigns weight one to the main treated unit, which happens, for example, when there are no pure controls. More generally, when $\widehat{w}_{2}\widehat{l}_{1}$ is close to one the denominator in \eqref{eq:tiscm} is small, and the iSCM estimator becomes sensitive to small discrepancies in the underlying fits: even when the effects are identified, the stability of the estimator depends on the magnitude of the cross-weights.

\subsection{The simplified iSCM}\label{sec:simplified}

A useful variant imposes $\widehat{l}_{1}=0$ by construction, that is, it estimates unit 2's synthetic control on the pure controls only; let $\widehat{\gamma}^{\,res}_{t}$ denote the gap of unit 2 computed from this restricted fit. With the approximation errors omitted, as in the system of Section~\ref{sec:mainresult}, the system then becomes triangular,
\begin{equation}\label{eq:simp}
\widehat{\theta}_{t}=\theta_{t}-\widehat{w}_{2}\gamma_{t},
\qquad
\widehat{\gamma}^{\,res}_{t}=\gamma_{t},
\end{equation}
no inversion is required, and the potential instability associated with the denominator in \eqref{eq:tiscm} does not arise. We refer to the resulting estimator,
\begin{equation}\label{eq:siscm}
\widehat{\theta}^{\,siSCM}_{t}\equiv\widehat{\theta}_{t}+\widehat{w}_{2}\,\widehat{\gamma}^{\,res}_{t},
\end{equation}
as the \textit{simplified} iSCM. Corollary~\ref{cor:plugin} below shows that the simplified iSCM replaces unit 2 in the synthetic control of the main treated unit with unit 2's own synthetic control based on pure controls, and Remark~\ref{rem:polarcases}(b) gives the general-$m$ analogue: for $m>2$, the simplified iSCM estimates the synthetic control of every unit in $M$ on the pure controls only, so that rows $2,\ldots,m$ of $\widehat{\Omega}$ reduce to unit vectors, the system is triangular, and the estimator is $\widehat{\theta}^{\,siSCM}_{t}=\widehat{\theta}_{t}+\sum_{j\in M}\widehat{w}_{j}\widehat{\gamma}^{\,res}_{jt}$, with estimation error $-B^{1}_{sc}-\sum_{j\in M}\widehat{w}_{j}\widetilde{B}^{j}_{sc}$, where $\widetilde{B}^{j}_{sc}$ denotes the approximation error of unit $j$'s restricted fit. The simplified iSCM thus avoids the inversion step, and hence the amplification factor, but it replaces the approximation error of each unrestricted fit with that of the corresponding restricted fit, which may be larger when the main treated unit is important for the fit of the units in $M$; Section~\ref{sec:biascomparison} quantifies this trade-off. The simplified iSCM is attractive when unit 2's fit does not deteriorate appreciably once the main treated unit is excluded from its donor pool and, as we argue in Sections \ref{sec:biascomparison} and \ref{sec:guidance}, it is the more reliable choice when the cross-weights are large, provided the restricted fits remain adequate.

\subsection{The implied pure-control weights}

Let the unrestricted synthetic control for the main treated unit place weight $\widehat{w}_{2}$ on the potentially affected unit and weights $\widehat{W}_{C}$ on the pure controls, and let the unrestricted synthetic control for unit 2 place weight $\widehat{l}_{1}$ on the main treated unit and weights $\widehat{L}_{C}$ on the pure controls (in the notation of Section~\ref{sec:purecontrol}, $\widehat{W}_{C}=\widehat{p}_{1}$ and $\widehat{L}_{C}=\widehat{p}_{2}$):
\[
\widehat{Y}_{1t}^{N}=\widehat{w}_{2}Y_{2t}+Y_{Ct}'\widehat{W}_{C},
\qquad
\widehat{Y}_{2t}^{N}=\widehat{l}_{1}Y_{1t}+Y_{Ct}'\widehat{L}_{C},
\]
with $\mathbf{1}_{C}'\widehat{W}_{C}=1-\widehat{w}_{2}$ and $\mathbf{1}_{C}'\widehat{L}_{C}=1-\widehat{l}_{1}$ under adding-up constraints.

\begin{proposition}\label{prop:implicit}
If $1-\widehat{w}_{2}\widehat{l}_{1}\neq 0$, the iSCM estimator for the main treated unit can be written as
\begin{equation}\label{eq:implicitrestricted}
\widehat{\theta}^{\,iSCM}_{t}=Y_{1t}-Y_{Ct}'\widehat{W}^{\,*}_{C},
\qquad
\widehat{W}^{\,*}_{C}\equiv\frac{\widehat{W}_{C}+\widehat{w}_{2}\widehat{L}_{C}}{1-\widehat{w}_{2}\widehat{l}_{1}}.
\end{equation}
\end{proposition}

\begin{corollary}\label{cor:convexity}
If the underlying SC-type estimator enforces nonnegativity and adding-up constraints, so that $\widehat{w}_{2},\widehat{l}_{1}\in[0,1]$, $\widehat{W}_{C}\ge 0$ and $\widehat{L}_{C}\ge 0$, and if $\widehat{w}_{2}\widehat{l}_{1}<1$, which holds whenever each fit places strictly positive total weight on the pure controls (Lemma~A.1), then $\widehat{W}^{\,*}_{C}\ge 0$ and $\mathbf{1}_{C}'\widehat{W}^{\,*}_{C}=1$, so \eqref{eq:implicitrestricted} is a convex combination of pure-control outcomes.
\end{corollary}

\begin{corollary}\label{cor:plugin}
If $\widehat{l}_{1}=0$, then $\widehat{W}^{\,*}_{C}=\widehat{W}_{C}+\widehat{w}_{2}\widehat{L}_{C}$, i.e.\ the simplified iSCM replaces the potentially affected unit in the synthetic control of the main treated unit with that unit's own synthetic control.
\end{corollary}

The proofs are in Appendix~A. Proposition~\ref{prop:implicit} is the $m=2$ counterpart of Lemma~\ref{lem:purecontrol} and makes the discussion of Section~\ref{sec:relationrestricted} concrete: although iSCM uses only pure-control outcomes \textit{ex post}, the implied weights $\widehat{W}^{\,*}_{C}$ are constructed from the two unrestricted fits and are generally not the solution to the restricted SCM problem that excludes unit 2 from the donor pool. Because the SCM objective and its constraints, or the regularization of other SC-type estimators, define the selected weights in a nonlinear way, the restricted optimizer can differ substantially from $\widehat{W}^{\,*}_{C}$, leading to a different pre-intervention fit and different finite-sample behaviour.

\subsection{Bias comparison}\label{sec:biascomparison}

We now reinstate the approximation errors $B^{1}_{sc}$ and $B^{2}_{sc}$ and compare the biases of the unrestricted SCM, the full and simplified iSCM, and the restricted SCM estimators of $\theta_{t}$. Throughout this subsection, $\mathrm{Bias}(\cdot)$ denotes the realized estimation error conditional on the estimated weights, in line with the terminology introduced in Section~\ref{sec:decomposition}; no expectation is taken. From \eqref{eq:dec1}, the bias of the unrestricted estimator $\widehat{\theta}_{t}$ is
\begin{equation}\label{eq:biasunres}
\mathrm{Bias}\big(\widehat{\theta}_{t}\big)=-B^{1}_{sc}-\widehat{w}_{2}\gamma_{t}.
\end{equation}
Specializing Theorem~\ref{thm:main}(a) to $m=2$, or substituting the biased estimates into \eqref{eq:tiscm} directly, yields
\begin{equation}\label{eq:biasiscm}
\mathrm{Bias}\big(\widehat{\theta}^{\,iSCM}_{t}\big)
=\frac{-B^{1}_{sc}-\widehat{w}_{2}B^{2}_{sc}}{1-\widehat{w}_{2}\widehat{l}_{1}}
=-\alpha\,B^{1}_{sc}-\alpha\,\widehat{w}_{2}B^{2}_{sc},
\qquad
\alpha\equiv\frac{1}{1-\widehat{w}_{2}\widehat{l}_{1}},
\end{equation}
where $\alpha\ge 1$ whenever $\widehat{w}_{2}\widehat{l}_{1}\in[0,1)$, as under nonnegativity and adding-up constraints.
Comparing \eqref{eq:biasunres} and \eqref{eq:biasiscm} term by term clarifies the trade-off. The first term of \eqref{eq:biasiscm} is larger in magnitude than that of \eqref{eq:biasunres} by the amplification factor $\alpha$, which is directly computable and should always be reported; in our application, where $\widehat{w}_{2}=0.42$ and $\widehat{l}_{1}=0.33$, $\alpha\approx 1.16$, so the amplification is about 16\%. The second terms are driven by $B^{2}_{sc}$ and $\gamma_{t}$, respectively. Since $B^{2}_{sc}$ is the approximation error in estimating unit 2's counterfactual, it is expected to be small relative to the effect $\gamma_{t}$ itself unless the chosen estimator performs poorly for unit 2. The full iSCM therefore removes the contamination bias $-\widehat{w}_{2}\gamma_{t}$ at the cost of amplifying the approximation errors by the factor $\alpha$. This cost is modest when $\widehat{w}_{2}\widehat{l}_{1}$ is far from one, but it grows without bound as $\widehat{w}_{2}\widehat{l}_{1}$ approaches one: when $\alpha$ is large, even small approximation errors in the underlying fits are magnified, and the correction can become unreliable precisely in the situations in which the two units are most intertwined.

The same calculation for the simplified iSCM in \eqref{eq:siscm} gives
\begin{equation}\label{eq:biassiscm}
\mathrm{Bias}\big(\widehat{\theta}^{\,siSCM}_{t}\big)=-B^{1}_{sc}-\widehat{w}_{2}\widetilde{B}^{2}_{sc},
\end{equation}
where $\widetilde{B}^{2}_{sc}$ denotes the approximation error of unit 2's restricted fit. No amplification occurs, but $\widetilde{B}^{2}_{sc}$ replaces $B^{2}_{sc}$ and may be larger when the main treated unit is important for unit 2's fit. The simplified iSCM thus trades a possibly less accurate fit for unit 2 against full protection from the instability associated with $\alpha$.

Now let $B_{rSC}$ denote the approximation error of the restricted SCM, so that its bias is $-B_{rSC}$. The bias difference between the restricted SCM and the full iSCM has the sign of $\big(B^{1}_{sc}+\widehat{w}_{2}B^{2}_{sc}\big)-\big(1-\widehat{w}_{2}\widehat{l}_{1}\big)B_{rSC}$, and the full iSCM has the smaller bias in magnitude if and only if $\big|B^{1}_{sc}+\widehat{w}_{2}B^{2}_{sc}\big|<\big(1-\widehat{w}_{2}\widehat{l}_{1}\big)\big|B_{rSC}\big|$, a comparison of unobserved approximation errors, so no method dominates \textit{a priori}. The comparison is nevertheless informative: $B^{1}_{sc}$ and $B^{2}_{sc}$ are the errors of two unrestricted fits, which can be made small when the donor pools are rich, whereas $B_{rSC}$ is the error of a fit that excludes precisely the units most similar to the main treated one and can be large when those units are essential, as in our application, where the restricted pre-intervention RMSPE is more than twice the unrestricted one. Conversely, when the restricted fit is comparable to the unrestricted one, the restricted SCM avoids the amplification factor $\alpha$ altogether and may be preferable; and when the affected units are essential for the fit but $\alpha$ is large, the simplified iSCM offers a middle course, removing the contamination without the inversion step. Section~\ref{sec:guidance} turns these comparisons into a practical decision rule.

\section{Empirical guidance}\label{sec:guidance}
\setcounter{equation}{0}
\setcounter{remark}{0}

\subsection{Choosing among the restricted SCM and the full and simplified iSCM}\label{sec:choosing}

The bias comparison of Section~\ref{sec:biascomparison} involves unobserved approximation errors, so the choice among the restricted SCM and the full and simplified iSCM must rest on observable diagnostics. In this section we organize these diagnostics into an explicit decision rule. Units that are potentially affected by the intervention are often precisely the units whose characteristics are most similar to those of the main treated unit, for example because they are geographically or economically close to it; indeed, \citet{Aba2020} suggests including units potentially affected by spillovers in the donor pool even though they induce a bias, which iSCM removes. We recommend estimating all four specifications, namely the unrestricted SCM, the restricted SCM, and the full and simplified iSCM, and choosing the main specification according to the following steps; the results of all specifications should be reported in any case, for the reasons discussed at the end of this subsection.

Recall that $X_{1}$ denotes the $(k\times 1)$ vector of predictors of the main treated unit and $X_{0}\equiv(X_{2},\ldots,X_{J})$ the $(k\times(J-1))$ matrix of donor predictors in the unrestricted pool, and let $X^{res}_{0}\equiv(X_{m+1},\ldots,X_{J})$ denote the corresponding matrix for the restricted pool, with associated estimated weights $\widehat{W}^{unres}$ and $\widehat{W}^{res}$. For a generic unit $i$ with weight vector $\widehat{W}$ on the donors $j\neq i$, define the pre-intervention RMSPE as
\begin{equation}\label{eq:rmspe}
\mathrm{RMSPE}_{i}\big(\widehat{W}\big)=\bigg(\frac{1}{T_{0}}\sum_{t=1}^{T_{0}}\Big(Y_{it}-\sum_{j\neq i}\widehat{w}_{j}Y_{jt}\Big)^{2}\bigg)^{1/2},
\end{equation}
so that the same definition applies to the main treated unit and to every unit in $M$.

\setcounter{bean}{0}
\begin{center}
\begin{list}
{\textsc{Step} \arabic{bean}.}{\usecounter{bean}}
\item Check whether the potentially affected units receive substantial weights in the unrestricted fit. If they do not, all specifications are expected to perform similarly (Remark~\ref{rem:smallweights}), and the restricted SCM can be adopted directly.
\item Compare the predictor imbalances of the two specifications, both predictor by predictor, as in Table~\ref{tab:predictors}, and through a scalar summary. As a default, and as implemented in our accompanying R package, we aggregate the absolute discrepancies into their sum; when the predictors are measured on very different scales, they should be standardized before aggregation.
\item Compare $\mathrm{RMSPE}_{1}(\widehat{W}^{unres})$ and $\mathrm{RMSPE}_{1}(\widehat{W}^{res})$.\footnote{Note that, unlike in a regression model, the restricted SCM may well achieve a \textit{lower} pre-intervention RMSPE than the unrestricted one: SC-type estimators do not minimize the outcome RMSPE in \eqref{eq:rmspe} directly. The original SCM, for example, minimizes the predictor imbalance $\|X_{1}-X_{0}W\|_{V}$, with the weighting matrix $V$ itself re-estimated for each donor pool, so the outcome RMSPE is not the optimized objective and need not improve when donors are added (see \citealp{Aba2020}). This is what makes the comparison in Steps 2 and 3 informative.} To evaluate the two fits, researchers can choose between a quick heuristic and a more rigorous validation process. As a fast, albeit somewhat arbitrary starting point, one could regard the fits as comparable when their RMSPEs differ by less than 25\%, and consider the restricted fit substantially worse if its RMSPE exceeds the unrestricted one by more than 50\%. Alternatively, for a less arbitrary but more time-consuming approach, one can use out-of-sample validation. This involves splitting the pre-intervention period into a training period (the first two-thirds of the observations) and a validation period (the final third), estimating both specifications on the training data, and comparing their RMSPEs over the validation period.
\item Compute the amplification implied by the cross-weights: for $m=2$, the factor $\alpha=1/(1-\widehat{w}_{2}\widehat{l}_{1})$ of \eqref{eq:biasiscm}; in general, $\sigma_{\min}(\widehat{\Omega})$, whose inverse bounds the propagation of approximation errors by Theorem~\ref{thm:main}(a). In addition, compare, for each $i\in M$, the pre-intervention RMSPEs of unit $i$'s unrestricted and restricted fits, since the simplified iSCM relies on the restricted fits of the units in $M$.
\end{list}
\end{center}

If the imbalances and the RMSPEs of the two specifications are similar, the restricted SCM is preferable, as it avoids the amplification of approximation errors discussed in Section~\ref{sec:biascomparison}. If the restricted specification displays substantially larger imbalances and/or a substantially larger RMSPE, the potentially affected units carry information that the restricted optimizer cannot exploit, and the iSCM correction is warranted. In that case, Step 4 guides the choice between the full and the simplified iSCM. When the amplification factor is close to one, say below 1.25, the full iSCM can be applied safely. When it is large, small approximation errors in the underlying fits are magnified, and the simplified iSCM is the more reliable choice provided that the restricted fits of the units in $M$ remain adequate; when the amplification factor is large and those restricted fits are also poor, no variant can be considered reliable, and the composition of $M$ should be reconsidered along the lines of Section~\ref{sec:choosingM}. When Steps 2 and 3 point in different directions, we give precedence to the predictor imbalance comparison and to the validation-split variant of Step 3, as the in-sample RMSPE comparison is more exposed to overfitting; this precedence rule is part of the procedure and is the one we apply in Section~\ref{sec:application}.

Steps 1--3 should also be repeated for each potentially affected unit, treating it as if it were the main treated unit, both to assess the plausibility of Assumption~\ref{ass:fitaffected} and to determine whether the main treated unit is needed in its donor pool. When it is not, the simplified iSCM can be used at no cost in fit, as in \eqref{eq:siscm}; more generally, the simplified iSCM remains available as an alternative whenever Step 4 signals a poorly conditioned system. Finally, it is worth comparing the results of iSCM and restricted SCM with those of the unrestricted SCM, bearing in mind that the latter is biased whenever a potentially affected unit receiving non-negligible weight is in fact affected.

The procedure just described selects a specification on the basis of the data, and the placebo analysis of Section~\ref{sec:inference} is typically conducted conditional on the selected specification; this conditioning is not accounted for and gives rise to a pre-test problem (\citealp{JR22}). For this reason, reporting the results of all four specifications is our default recommendation rather than an optional robustness check, and the decision rule should be read as selecting the specification to be emphasized, not as a licence to discard the others.

Whatever the selected specification, we recommend reporting: the $m$ estimated weight vectors; the matrix $\widehat{\Omega}$, together with $\sigma_{\min}(\widehat{\Omega})$ and, for $m=2$, the amplification factor $\alpha$; the pre-intervention RMSPEs of the unrestricted and restricted fits of the main treated unit and of every unit in $M$; and the estimated effects under all four specifications. All of these objects are already produced by Steps 1--4, so the additional reporting burden is negligible.

\subsection{Choosing the set of potentially affected units}\label{sec:choosingM}

In the presence of spillovers, iSCM requires the researcher to specify \textit{ex ante} which units may be indirectly affected by the intervention. This choice determines both the extent of the contamination bias and the stability of the correction. We recommend combining two criteria. First, rely on economic and institutional considerations, such as geographical proximity, historical ties, supply-chain linkages or trade integration, to identify a plausible set of potentially affected units, in line with the spillover literature, which typically restricts interactions to pre-specified groups (\citealp{Cer2017}, \citealp{For2018}, \citealp{Hub2019}, and \citealp{Vaz2017}). Second, focus on units that receive non-negligible weights in the unrestricted fit, as these are the only units capable of transmitting a non-negligible contamination bias (Remark~\ref{rem:smallweights}).

At the same time, classifying too many units as potentially affected may deteriorate the conditioning of $\widehat{\Omega}$ and amplify estimation errors, as discussed in Sections \ref{sec:mainresult} and \ref{sec:m2}. The choice of $M$ should therefore balance these considerations, including units that are both economically exposed and empirically influential, while avoiding an unnecessarily large or weakly connected set. In our application, Austria satisfies both criteria: it receives the largest weight in synthetic West Germany and has strong historical and economic ties with Germany. By contrast, the United States and Japan, which also receive sizeable weights, are economically much less exposed to reunification, while France, arguably exposed through trade, receives zero weight, so any spillover on France cannot contaminate the estimate (Remark~\ref{rem:smallweights}). A simple and transparent sensitivity analysis consists of comparing the iSCM estimates obtained under a minimal specification of $M$, including only the units most clearly exposed, with those obtained under an extended specification including all plausibly exposed units.

\section{Inference}\label{sec:inference}
\setcounter{equation}{0}

Inference for synthetic control methods is complicated by the small number of units, and the most common approach in applied work is the placebo, or permutation, analysis of \citet{Aba2010} and \citet{Aba2015}; see \citet{Che2019}, \citet{Fir2018} and \citet{Li2019} for formal procedures. In our framework an additional complication arises: the observed post-intervention outcomes of the potentially affected units embed the effects of the intervention, and test statistics must therefore be computed from outcomes that have been purged of the estimated effects. With this modification, existing procedures can be adapted to our setting. We describe the adaptation of the in-space placebo analysis in detail; we view the resulting procedure as a transparent diagnostic, in the spirit of the original placebo studies, rather than as a test with formal size guarantees. Note that the comparison is meaningful precisely because the approximation errors $B^{j}_{sc}$ are not exactly zero in finite samples: by Theorem~\ref{thm:main}(a), the corrected gaps reflect finite-sample fitting errors, and the placebo analysis assesses whether the gap of the unit of interest is large relative to the distribution of such errors across units.

For each unit we define a scalar statistic given by the ratio of the post-intervention to the pre-intervention RMSPE, computed on corrected outcomes. When testing the effect on the main treated unit, the post-intervention outcomes of all potentially affected units are corrected by subtracting their estimated iSCM effects, yielding the corrected synthetic outcome
\begin{equation}\label{eq:tildemain}
\widetilde{Y}^{N}_{1t}\equiv\sum_{j=m+1}^{J}\widehat{w}_{j}Y_{jt}+\sum_{j=2}^{m}\widehat{w}_{j}\big(Y_{jt}-\widehat{\gamma}^{\,iSCM}_{jt}\big),
\quad t>T_{0},
\end{equation}
and the statistic
\begin{equation}\label{eq:r1}
r_{1}=\frac{\Big(\frac{1}{T-T_{0}}\sum_{t=T_{0}+1}^{T}\big(Y_{1t}-\widetilde{Y}^{N}_{1t}\big)^{2}\Big)^{1/2}}{\Big(\frac{1}{T_{0}}\sum_{t=1}^{T_{0}}\big(Y_{1t}-\widehat{Y}^{N}_{1t}\big)^{2}\Big)^{1/2}}.
\end{equation}
Since no unit is affected before the intervention, corrected and uncorrected outcomes coincide for $t\le T_{0}$, so the denominator of \eqref{eq:r1} is unambiguous. Moreover, the numerator has a convenient interpretation: the first row of the identity $\widehat{\beta}_{t}=\widehat{\Omega}\,\widehat{\vartheta}^{\,iSCM}_{t}$ gives
\[
Y_{1t}-\widetilde{Y}^{N}_{1t}=\widehat{\theta}_{1t}+\sum_{j=2}^{m}\widehat{w}_{j}\widehat{\gamma}^{\,iSCM}_{jt}=\widehat{\theta}^{\,iSCM}_{1t},
\]
so the corrected gap coincides with the iSCM estimate, and $r_{1}$ compares the post-intervention RMSPE of the iSCM gaps with the pre-intervention RMSPE of the unrestricted fit. Similarly, when testing the effect on a potentially affected unit $i\in M$, the outcomes of the main treated unit and of all other potentially affected units are corrected, leaving unit $i$'s outcome untouched:
\begin{equation}\label{eq:tildei}
\widetilde{Y}^{N}_{it}\equiv\sum_{j=m+1}^{J}\widehat{l}^{\,i}_{j}Y_{jt}+\sum_{j\in M\setminus\{i\}}\widehat{l}^{\,i}_{j}\big(Y_{jt}-\widehat{\gamma}^{\,iSCM}_{jt}\big)+\widehat{l}^{\,i}_{1}\big(Y_{1t}-\widehat{\theta}^{\,iSCM}_{1t}\big),
\quad t>T_{0},
\end{equation}
and
\begin{equation}\label{eq:ri}
r_{i}=\frac{\Big(\frac{1}{T-T_{0}}\sum_{t=T_{0}+1}^{T}\big(Y_{it}-\widetilde{Y}^{N}_{it}\big)^{2}\Big)^{1/2}}{\Big(\frac{1}{T_{0}}\sum_{t=1}^{T_{0}}\big(Y_{it}-\widehat{Y}^{N}_{it}\big)^{2}\Big)^{1/2}}.
\end{equation}
By the corresponding row of the same identity, $Y_{it}-\widetilde{Y}^{N}_{it}=\widehat{\gamma}^{\,iSCM}_{it}$, so the numerator of \eqref{eq:ri} is the post-intervention RMSPE of unit $i$'s iSCM gaps.

The placebo procedure is then carried out separately for the main treated unit and for each potentially affected unit, as follows.

\setcounter{bean}{0}
\begin{center}
\begin{list}
{\textsc{Step} \arabic{bean}.}{\usecounter{bean}}
\item Compute $r_{1}$ in \eqref{eq:r1}, or $r_{i}$ in \eqref{eq:ri}, for the unit whose effect is being tested.
\item For each pure-control unit $j\in C$, reassign the intervention to unit $j$, which plays the role of the unit being tested. Using pre-intervention data only, estimate the SC-type weights for unit $j$ with all remaining units in its donor pool. For $t>T_{0}$, compute unit $j$'s placebo gap on corrected outcomes, subtracting from the outcomes of the main treated unit and of each unit in $M$ their iSCM effects $\widehat{\theta}^{\,iSCM}_{1t}$ and $\widehat{\gamma}^{\,iSCM}_{it}$, estimated once from the original system of Section~\ref{sec:implementation}, since these units are genuinely affected regardless of the unit being tested. The set $M$ is held fixed across placebo runs, no effect is estimated for unit $j$ itself, and the denominator of $r_{j}$ is the pre-intervention RMSPE of unit $j$'s fit.
\item Compare $r_{1}$, or $r_{i}$, with the empirical distribution of the placebo statistics $\{r_{j}: j\in C\}$; the rank of the statistic in this distribution can be reported as a placebo $p$-value.
\end{list}
\end{center}

Two features of the procedure should be borne in mind when interpreting the ranking. First, the corrected outcomes entering all statistics embed the estimation error of the purged effects, a common component that induces dependence across the placebo statistics. Second, by the identity above, the numerator of $r_{1}$ is the post-intervention RMSPE of the iSCM gaps, which amplifies the approximation errors of the underlying fits by the factor $\alpha$ of Section~\ref{sec:biascomparison}, whereas the placebo gaps of pure controls that are weakly connected to the units in $\{1\}\cup M$ embed no comparable amplification; under the null hypothesis of no effect, $r_{1}$ is therefore mechanically tilted upwards relative to such placebos, and the resulting $p$-values err on the side of rejection. When $\alpha$ is close to one, as in our application, this distortion is small; for large $\alpha$, the placebo analysis inherits the instability of the full correction, and we recommend reporting $\alpha$, or $\sigma_{\min}(\widehat{\Omega})$, alongside the placebo statistics. In addition, with $|C|$ pure controls the attainable placebo $p$-values are multiples of $1/(|C|+1)$; in our application $|C|=15$, so the smallest attainable $p$-value is $1/16\approx 0.06$, and conventional 5\% statements are unattainable.

A further caveat is that the placebo analysis is typically conducted conditional on the specification selected by the procedure of Section~\ref{sec:choosing}; as discussed there, this conditioning is not accounted for, which reinforces the recommendation to report the results of all specifications (see \citealp{JR22}).

The same idea applies to other inference procedures available in the literature (e.g., \citealp{Cao2019}; \citealp{Che2019}; \citealp{Fir2018}; \citealp{Gob2016}; \citealp{Li2019}). For example, the conformal procedure of \citet{Che2019} can be adapted to our framework by accounting for the fact that the observed post-intervention outcome of each potentially affected unit includes its treatment effect, both in the construction of the data under the null hypothesis and in the computation of the residuals.

\section{Empirical application: German reunification}\label{sec:application}
\setcounter{equation}{0}

In this section we use iSCM to estimate the effect of German reunification on West Germany's GDP per capita, allowing for spillover effects on Austria. In October 1990, less than a year after the fall of the Berlin Wall, the German Democratic Republic (East Germany) and the Federal Republic of Germany (West Germany) were officially reunified. This event, one of the most important milestones of post-war European history, most likely affected not only the German economy but also the economies of other European countries.

As discussed in \citet{Aba2015} and \citet{Aba2019}, reunification could have had negative spillover effects on Austria's economic growth, as West Germany diverted demand and investment from Austria towards East Germany. Austria has historically had close links with Germany: the two countries share the same language and, to a great extent, a common history, and their economic cooperation continued throughout the Cold War. These close cultural and economic ties make Austria a plausible recipient of spillover effects from German reunification, rendering it potentially unsuitable as a donor under the standard no-interference assumption. At the same time, Austria provides valuable pre-intervention information for constructing a synthetic West Germany. As shown below, it receives the largest synthetic control weight, and excluding it substantially worsens the pre-intervention fit. This trade-off makes this empirical setting particularly well-suited for the iSCM. The method allows Austria to remain in the donor pool while correcting for potential contamination, jointly estimating the direct effect of reunification on West Germany and its spillover effect on Austria.

We use the country-level panel data of \citet{Aba2015}, covering 1960--2003 with the post-intervention period starting in 1990. The sample comprises West Germany, Austria and 15 other countries of the Organisation for Economic Co-operation and Development (OECD): Australia, Belgium, Denmark, France, Greece, Italy, Japan, the Netherlands, New Zealand, Norway, Portugal, Spain, Switzerland, the United Kingdom and the United States. The outcome is real GDP per capita at purchasing power parity, measured in 2002 US dollars; Appendix~B records the data sources.

We first replicate the SCM estimates of \citet{Aba2015} and inspect the weight assigned to Austria.\footnote{We refer to \citet{Aba2015} for a detailed description of the specification and the estimation procedure.} Austria receives the largest weight, $\widehat{w}_{A}=0.42$, so if Austria is affected by spillovers the unrestricted SCM estimate is contaminated.\footnote{The remaining European countries receive very little weight, so spillover effects on them would induce a negligible bias, as discussed in Remark~\ref{rem:smallweights}; a sensitivity analysis treating Switzerland and the Netherlands as potentially affected units is reported at the end of this section.} Following Section~\ref{sec:guidance}, we implement all four specifications and apply the decision rule of Section~\ref{sec:choosing}. Table~\ref{tab:predictors} shows that the unrestricted synthetic West Germany (second column) is much closer to West Germany (first column) in terms of observable characteristics than the restricted version (third column). Because the predictors are measured on very different scales, we follow Step 2 of Section 5.1 and standardize each predictor by its cross-sectional standard deviation before aggregating the absolute discrepancies; the resulting scalar imbalance is 0.26 for the unrestricted specification against 1.50 for the restricted one. The pre-intervention RMSPE of the restricted specification (260.74) is more than twice that of the unrestricted one (119.07), and the validation-split variant of Step 3, which estimates the weights on 1960 to 1979, the first two thirds of the pre-intervention period, and evaluates them on 1980 to 1989, widens the gap further (358.09 against 93.08). All three comparisons indicate that the restricted optimizer does not deliver a comparable fit, so the information contained in Austria's trajectory is essential and the iSCM correction is warranted.

We then repeat the same diagnostics to decide whether West Germany should be included in synthetic Austria's donor pool. To construct synthetic Austria we depart from the specification used for West Germany in one respect, and only for Austria: we select the predictor weights with the data-driven procedure of \citet{Aba2010}, rather than with the sample-splitting method of \citet{Aba2015}, because the latter produces a poor pre-intervention fit for Austria. A plausible explanation is the episode documented by \citet{Geh2018}: in 1980, shortly before the sample-split cut-off, Austria provided several loans to East Germany and in return Austrian nationalized industries received large-scale orders, which most likely stimulated Austria's exports. This is a deliberate deviation, for one unit, from the uniform specification recommended in Section~\ref{sec:guidance}, and we report it explicitly. The choice uses pre-intervention information only, although we acknowledge that selecting the predictor-weighting procedure on the basis of the realized pre-intervention fit is itself a specification choice, which reinforces the recommendation of Section~\ref{sec:choosing} to report all specifications. As shown in Table~\ref{tab:weights}, synthetic Austria assigns the largest weight to West Germany, $\widehat{l}_{WG}=0.33$. The unrestricted synthetic Austria reproduces Austria's pre-reunification characteristics better than the restricted version for all predictors except GDP per capita (Table~\ref{tab:predictors}); standardizing the discrepancies as in Step 2, the scalar imbalance is 0.53 for the unrestricted specification against 2.06 for the restricted one, while the in-sample pre-intervention RMSPE of the restricted specification (181.22) is slightly lower than that of the unrestricted one (194.67). However, the difference, approximately 7\%, falls within the comparability range defined in Step 3 and therefore does not provide sufficient grounds for excluding West Germany from Austria’s donor pool. As an additional diagnostic of the stability of the two fits, we estimate the weights over 1960–1979 and evaluate them over the held-out period 1980–1989. In this validation exercise, the unrestricted specification performs substantially better, with an RMSPE of 206.44 compared with 387.36 for the restricted specification. Taken together, these results support retaining West Germany in Austria’s donor pool. We therefore adopt the full iSCM as our main specification, re-estimating its weights over the entire pre-intervention period to exploit all available pre-treatment information. The deterioration of Austria's restricted fit in the validation comparison also matters for Step 4: it indicates that the simplified iSCM, which relies on that restricted fit, is less attractive in this application, although we report its estimates below.

\begin{table}
\caption{\label{tab:predictors}Economic growth predictors before German reunification.}
\begin{center}
\begin{tabular*}{\textwidth}{@{\extracolsep{\fill}}lccccc}
\hline\hline
 & & Unrestricted & Restricted & Unrestricted & Restricted \\
West Germany & Observed & synthetic & synthetic & discrepancy & discrepancy \\
\hline
GDP per capita & 15,808.90 & 15,804.64 & 16,138.83 & 4.26 & 329.93 \\
Trade openness & 56.78 & 56.91 & 50.73 & 0.14 & 6.04 \\
Inflation rate & 2.60 & 3.51 & 3.38 & 0.91 & 0.79 \\
Industry share & 34.54 & 34.38 & 33.30 & 0.15 & 1.24 \\
Schooling & 55.50 & 55.23 & 50.71 & 0.27 & 4.79 \\
Investment rate & 27.02 & 27.04 & 25.70 & 0.02 & 1.31 \\
\hline
 & & Unrestricted & Restricted & Unrestricted & Restricted \\
Austria & Observed & synthetic & synthetic & discrepancy & discrepancy \\
\hline
GDP per capita & 10,781.80 & 10,798.41 & 10,778.61 & 16.61 & 3.19 \\
Trade openness & 69.45 & 69.43 & 83.13 & 0.02 & 13.68 \\
Inflation rate & 4.91 & 4.92 & 5.59 & 0.01 & 0.68 \\
Industry share & 37.81 & 37.81 & 37.58 & 0.00 & 0.23 \\
Schooling & 53.25 & 45.71 & 35.44 & 7.54 & 17.81 \\
Investment rate & 26.64 & 26.64 & 27.03 & 0.00 & 0.38 \\
\hline\hline
\end{tabular*}
\end{center}
\footnotesize
\renewcommand{\baselineskip}{11pt}
\textbf{Note:} The first column reports the observed pre-reunification predictor values for West Germany (upper panel) and Austria (lower panel). The second and third columns report the corresponding values of the synthetic unit constructed with the unrestricted donor pool and with the restricted donor pool, which excludes Austria (upper panel) and West Germany (lower panel), respectively. The discrepancy columns report the absolute differences between the observed and the synthetic values, computed from unrounded figures.
\end{table}

\begin{table}
\caption{\label{tab:weights}Synthetic control weights for West Germany and Austria.}
\begin{center}
\begin{tabular*}{\textwidth}{@{\extracolsep{\fill}}lcccc}
\hline\hline
 & \multicolumn{2}{c}{West Germany} & \multicolumn{2}{c}{Austria} \\
Country & Unrestricted & Restricted & Unrestricted & Restricted \\
\hline
West Germany & -- & -- & 0.33 & -- \\
Austria & 0.42 & -- & -- & -- \\
Australia & 0 & 0 & 0 & 0 \\
Belgium & 0 & 0 & 0.12 & 0.51 \\
Denmark & 0 & 0 & 0 & 0 \\
France & 0 & 0 & 0 & 0 \\
Greece & 0 & 0 & 0 & 0 \\
Italy & 0 & 0 & 0 & 0 \\
Japan & 0.16 & 0.22 & 0.21 & 0.31 \\
Netherlands & 0.09 & 0.30 & 0.31 & 0.06 \\
New Zealand & 0 & 0 & 0 & 0 \\
Norway & 0 & 0 & 0.03 & 0 \\
Portugal & 0 & 0 & 0 & 0 \\
Spain & 0 & 0 & 0 & 0 \\
Switzerland & 0.11 & 0.09 & 0 & 0.12 \\
United Kingdom & 0 & 0 & 0 & 0 \\
United States & 0.22 & 0.40 & 0 & 0 \\
\hline\hline
\end{tabular*}
\end{center}
\footnotesize
\renewcommand{\baselineskip}{11pt}
\textbf{Note:} Estimated synthetic control weights for West Germany and Austria. The unrestricted columns use the full donor pool; the restricted columns exclude Austria from West Germany's donor pool and West Germany from Austria's donor pool, respectively. Weights are rounded to two decimal places; because of rounding, the weights in a column may not sum exactly to one.
\end{table}

With $\widehat{w}_{A}=0.42$ and $\widehat{l}_{WG}=0.33$, the estimated weight matrix is
\begin{align*}
\widehat{\Omega}=
\begin{pmatrix}
1 & -0.42\\
-0.33 & 1
\end{pmatrix}.
\end{align*}
The realized system is well conditioned: $\sigma_{\min}(\widehat{\Omega})\approx 0.62$ and $\det(\widehat{\Omega})=0.86$, and both off-diagonal row sums (0.42 and 0.33) are well below one, so $\widehat{\Omega}$ is strictly diagonally dominant and nonsingular (Lemma~A.1), and the inversion step is stable. The amplification factor of Section~\ref{sec:biascomparison} is $\alpha\approx 1.16$, implying that the correction inflates the approximation errors by about 16\%, a modest cost; by Step 4 of Section~\ref{sec:choosing}, the full iSCM can therefore be applied safely in this application. These quantities are also the ingredients of the bound \eqref{eq:rmspebound}: the pre-intervention RMSPE of the implied pure-control weights is at most $1.16\times(119.07+0.42\times 194.67)\approx 233$, and its exact value is 138.54, well below the 260.74 attained by the restricted optimizer, as discussed in Section~\ref{sec:relationrestricted}. The implied weights load mainly on Japan (0.28), the United States (0.25) and the Netherlands (0.25). Using the unrestricted estimates $\widehat{\theta}_{t}$ and $\widehat{\gamma}_{t}$, the iSCM estimators in \eqref{eq:tiscm} become
\[
\widehat{\theta}^{\,iSCM}_{t}=1.16\big(\widehat{\theta}_{t}+0.42\,\widehat{\gamma}_{t}\big)
\qquad\text{and}\qquad
\widehat{\gamma}^{\,iSCM}_{t}=1.16\big(\widehat{\gamma}_{t}+0.33\,\widehat{\theta}_{t}\big).
\]

Figure~\ref{fig:main} reports our main results. Panels (a) and (b) show the estimated gaps for West Germany and Austria obtained with the unrestricted SCM, the restricted SCM and iSCM; panels (c) and (d) plot the corresponding GDP per capita trajectories over 1960--2003.\footnote{For $t\le T_{0}$ we plot the unrestricted gap for iSCM, since the contamination correction is defined only for $t>T_{0}$; this is a plotting convention, and the pre-intervention fit of the implied pure-control weights of Section~\ref{sec:purecontrol} is a distinct object (see Section~\ref{sec:relationrestricted}).} For West Germany, both the unrestricted and the inclusive synthetic versions reproduce West Germany's pre-reunification GDP per capita almost perfectly, whereas excluding Austria substantially worsens the pre-reunification fit, confirming the importance of keeping Austria in the donor pool. In their original study, \citet{Aba2015} find that reunification reduced West Germany's GDP per capita by about 7.7\% per year on average over the post-reunification period, relative to the 1990 baseline; our unrestricted SCM estimates replicate theirs. Table~\ref{tab:effects} reports the estimated effects at selected years, together with averages over 1990 to 2003, of all four specifications:
the unrestricted SCM, the restricted SCM, the full iSCM and the
simplified iSCM. The iSCM estimates confirm the conjecture of \citet{Aba2015} about the direction of the bias: the estimated effect becomes more negative once spillovers are accounted for, with an average effect of $-1{,}677$ US dollars per year against $-1{,}570$ for the unrestricted SCM, a difference of about 7\% on average that reaches about 15\% in individual years once the effect has become sizeable. The restricted SCM, by contrast, yields a substantially larger effect,
averaging $-2{,}387$ US dollars per year compared with $-1{,}677$ of
the full iSCM, a difference of about 42\% in magnitude. The annual
difference exceeds 1,000 US dollars by 2003. This discrepancy is driven
by the poor pre-intervention fit obtained when Austria is excluded from West Germany's donor pool. The simplified iSCM delivers a smaller average effect, $-1{,}394$ US dollars per year: it replaces Austria's unrestricted fit with the restricted one, which the validation comparison above flags as appreciably worse, so the decision rule of Section~\ref{sec:choosing} favours the full iSCM in this application.

\begin{table}
\caption{\label{tab:effects}Estimated effects of German reunification at selected years.}
\begin{center}
\begin{tabular*}{\textwidth}{@{\extracolsep{\fill}}lcccc}
\hline\hline
West Germany & Unrestricted & Restricted & Full iSCM & Simplified iSCM \\
\hline
1990 & 8 & $-$491 & $-$83 & $-$107 \\
1995 & $-$1,217 & $-$1,860 & $-$1,293 & $-$1,172 \\
2000 & $-$2,645 & $-$3,595 & $-$2,639 & $-$2,077 \\
2003 & $-$3,372 & $-$4,560 & $-$3,529 & $-$3,026 \\
Average 1990--2003 & $-$1,570 & $-$2,387 & $-$1,677 & $-$1,394 \\
\hline
Austria & Unrestricted & Restricted & Full iSCM & Simplified iSCM \\
\hline
1990 & $-$188 & $-$272 & $-$216 & $-$272 \\
1995 & 249 & 106 & $-$181 & 106 \\
2000 & 894 & 1,351 & 15 & 1,351 \\
2003 & 802 & 824 & $-$373 & 824 \\
Average 1990--2003 & 304 & 418 & $-$254 & 418 \\
\hline\hline
\end{tabular*}
\end{center}
\footnotesize
\renewcommand{\baselineskip}{11pt}
\textbf{Note:} Estimated effects of reunification on GDP per capita, in 2002 US
dollars, rounded to the nearest dollar. The upper panel reports
$\widehat{\theta}_{t}$ (unrestricted SCM),
$\widehat{\theta}^{\,res}_{t}$ (restricted SCM),
$\widehat{\theta}^{\,iSCM}_{t}$, and
$\widehat{\theta}^{\,siSCM}_{t}$ for West Germany; the lower panel reports the
corresponding estimates of the spillover effect on Austria. For Austria, the
simplified iSCM estimate coincides with the restricted SCM gap
$\widehat{\gamma}^{\,res}_{t}$, since the simplified variant estimates Austria's
synthetic control using only the pure controls.
\end{table}

For Austria, iSCM estimates a small negative spillover effect, negative in eleven of the fourteen post-reunification years, with a decrease in GDP per capita of up to 709 US dollars (in 2001) and of 254 US dollars per year on average (Table~\ref{tab:effects}); this effect, however, cannot be distinguished from placebo variation: Austria's post- to pre-reunification RMSPE ratio is the second lowest in its donor pool, corresponding to a placebo $p$-value of $15/16$ (see Figure~B.1 in Appendix~B). For West Germany, in contrast, the same ratio is by far the largest in its donor pool, corresponding to a placebo $p$-value of $1/16\approx 0.06$, the smallest value attainable with 15 pure controls (Section~\ref{sec:inference}). It is worth noting that both the unrestricted and the restricted SCM estimate \textit{positive} spillover effects on Austria, of up to 894 and 1,351 US dollars, respectively. However, we know from Lemma~\ref{lem:affected} that the unrestricted estimate is contaminated by $-\widehat{l}_{WG}\theta_{t}$, and since the evidence strongly indicates that $\theta_{t}$ is negative, the unrestricted SCM is expected to overestimate $\gamma_{t}$. The fact that both benchmarks deliver implausibly positive spillover estimates confirms the importance of including West Germany in Austria's donor pool and of applying the iSCM correction in this application.

Finally, we assess the sensitivity of the results to the composition of $M$, as recommended in Section~\ref{sec:choosingM}. Treating Switzerland and the Netherlands, the other two European countries that receive non-negligible weight in synthetic West Germany, as potentially affected in addition to Austria ($m=4$) leaves the conclusions unchanged: the average effect on West Germany becomes $-1{,}752$ US dollars per year, about 4.5\% larger in magnitude than under the minimal specification, $\widehat{\Omega}$ remains strictly diagonally dominant, and $\sigma_{\min}(\widehat{\Omega})$ falls from 0.62 to 0.59, so the amplification bound $1/\sigma_{\min}(\widehat{\Omega})$ rises from 1.60 to about 1.70.

\begin{figure}
\centering
\begin{tabular}{cc}
\includegraphics[width=0.46\textwidth]{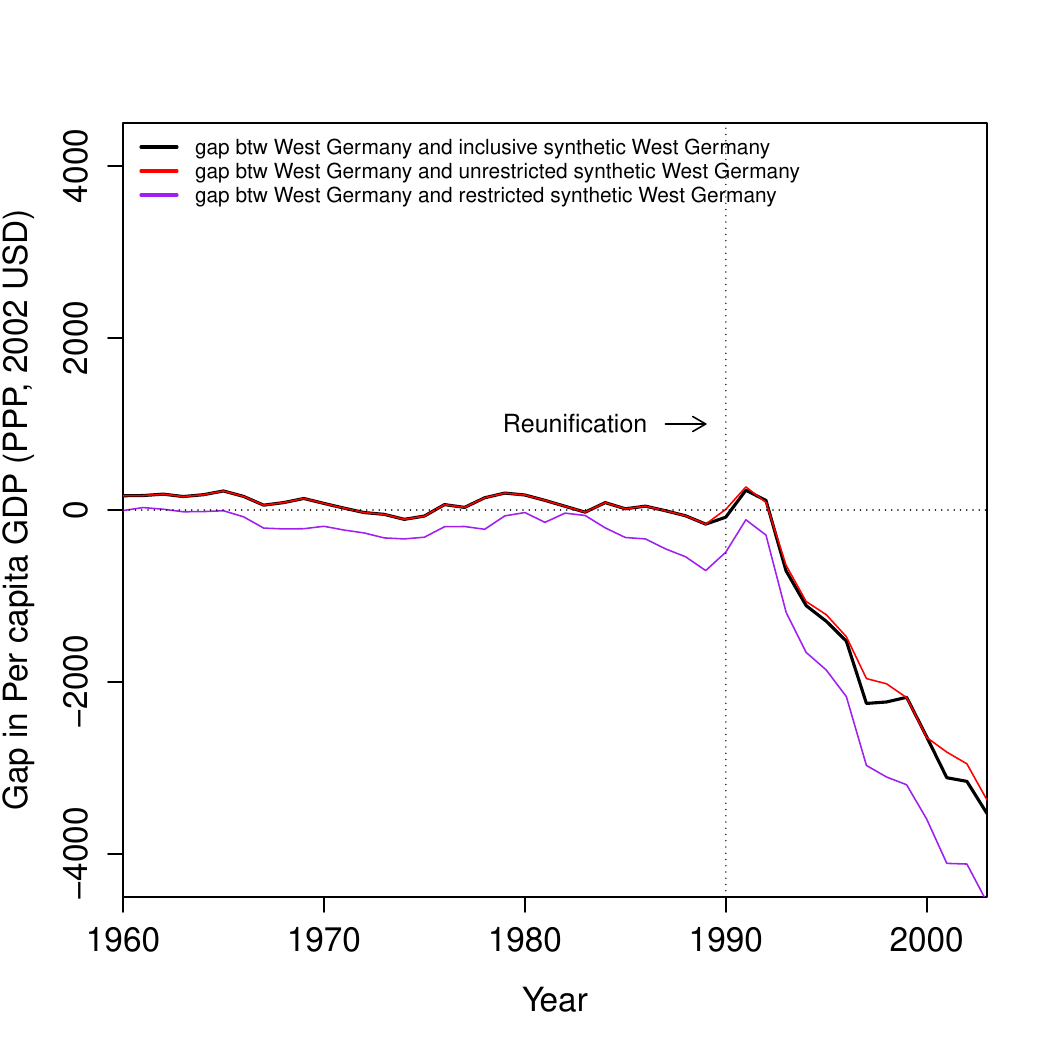} &
\includegraphics[width=0.46\textwidth]{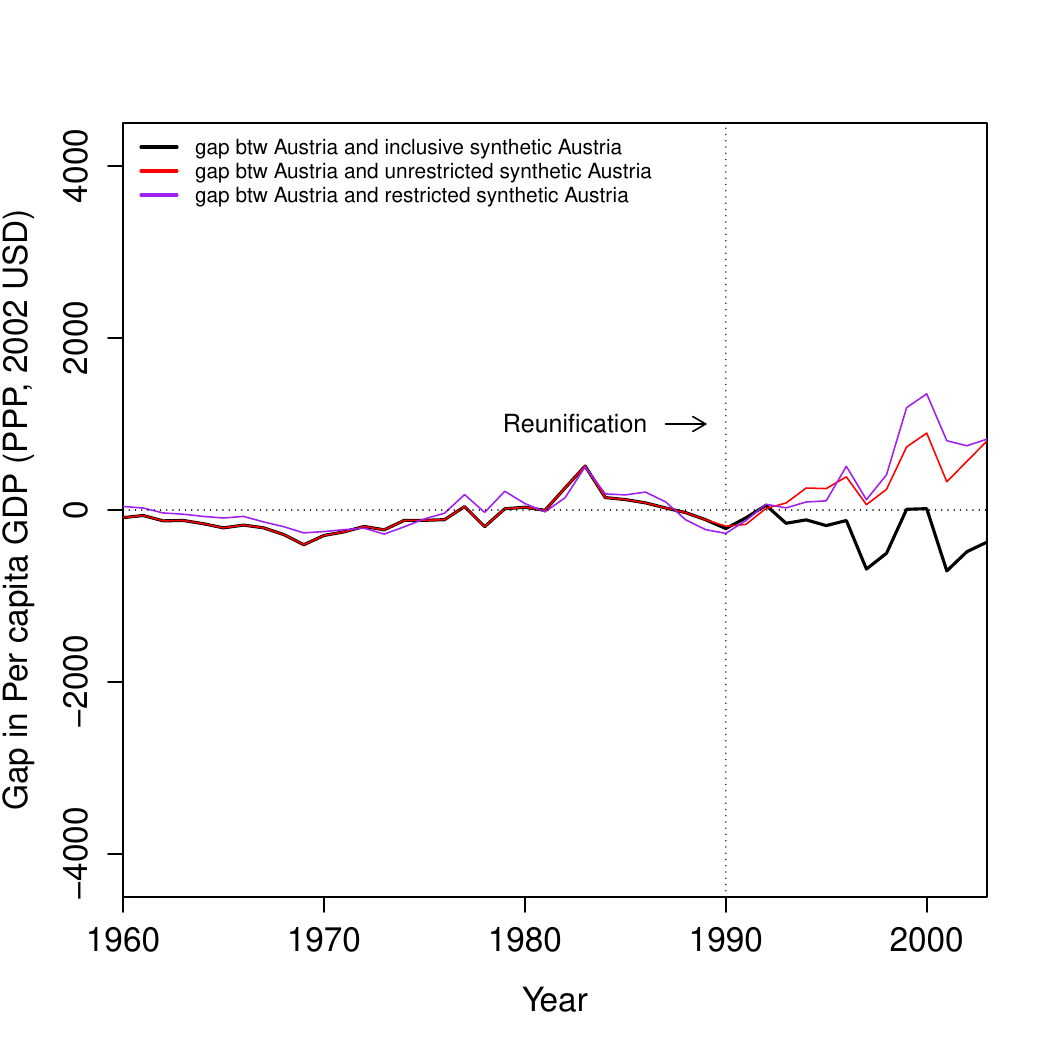} \\
{\footnotesize (a) Gaps, West Germany} & {\footnotesize (b) Gaps, Austria} \\[4pt]
\includegraphics[width=0.46\textwidth]{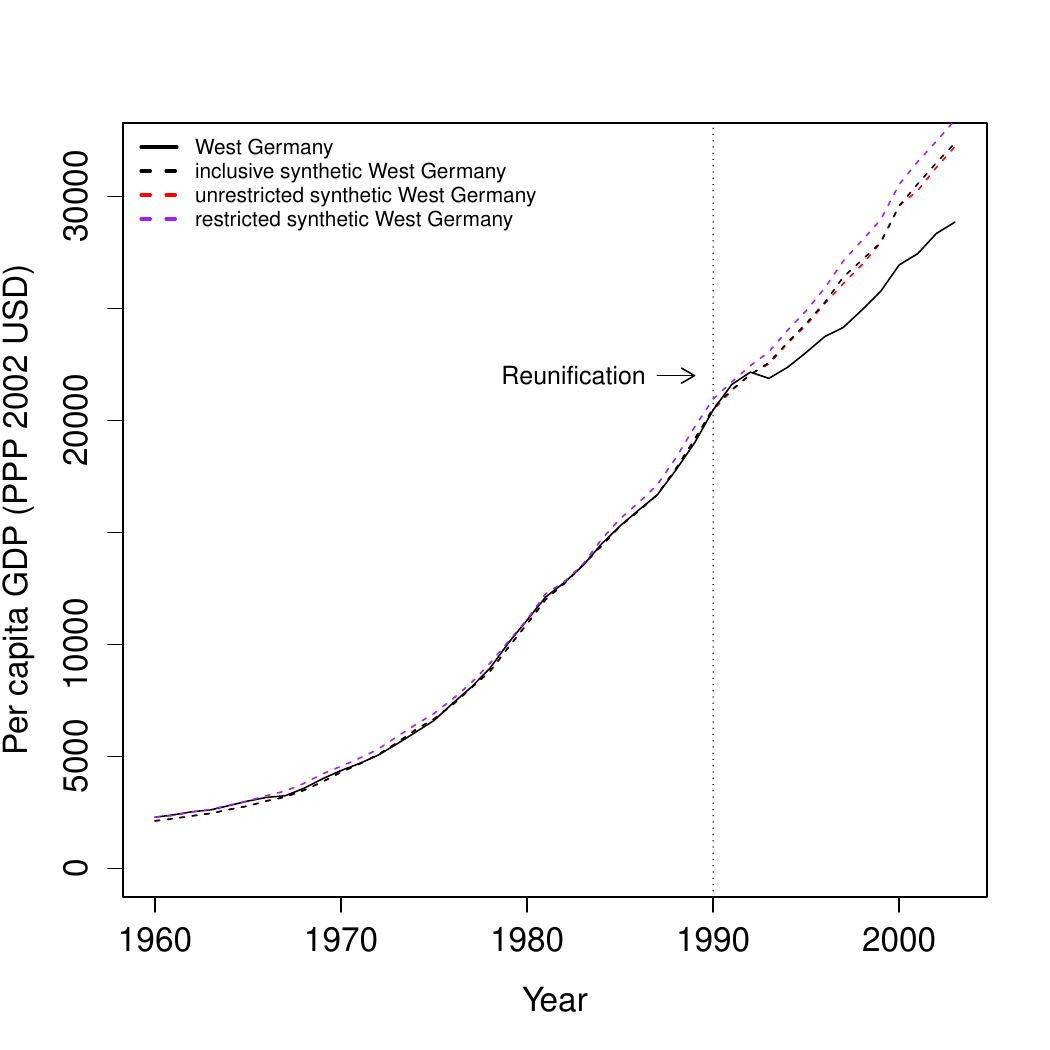} &
\includegraphics[width=0.46\textwidth]{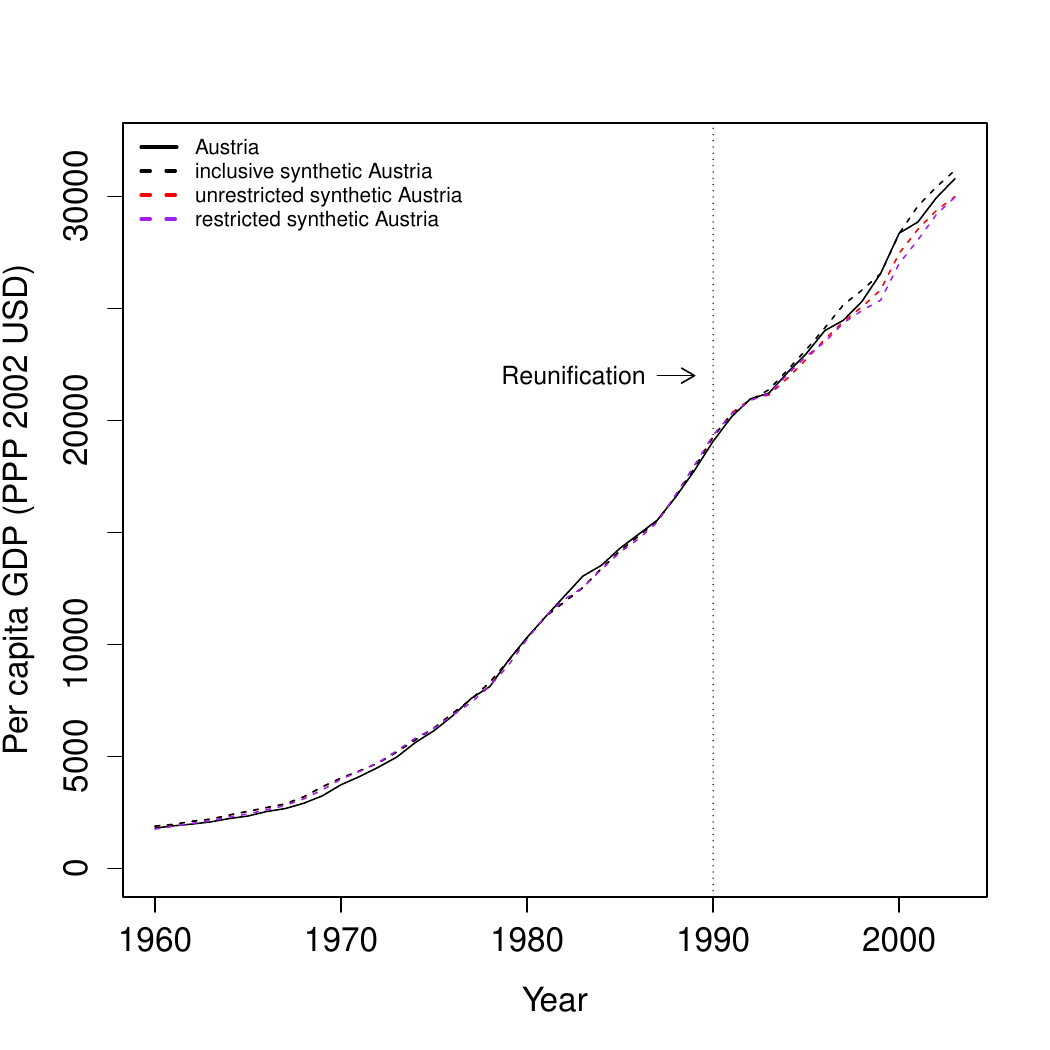} \\
{\footnotesize (c) Trends, West Germany} & {\footnotesize (d) Trends, Austria} \\
\end{tabular}
\caption{\label{fig:main}Estimated effects of German reunification.}
\footnotesize
\renewcommand{\baselineskip}{11pt}
\textbf{Note:} Panels (a) and (b) plot, for West Germany and Austria respectively, the gaps between the observed GDP per capita and the corresponding synthetic control under the unrestricted SCM, the restricted SCM, which excludes Austria from West Germany's donor pool and West Germany from Austria's, and iSCM, over 1960--2003; the vertical line marks reunification in 1990. Panels (c) and (d) plot the observed and synthetic GDP per capita trajectories. For $t\le T_{0}$ the iSCM gap coincides with the unrestricted one by the plotting convention discussed in Section~\ref{sec:relationrestricted}, as the correction operates only after the intervention.
\end{figure}

\subsection{The inclusive penalized SCM}\label{sec:pascm}

\begin{figure}
\centering
\begin{tabular}{cc}
\includegraphics[width=0.46\textwidth]{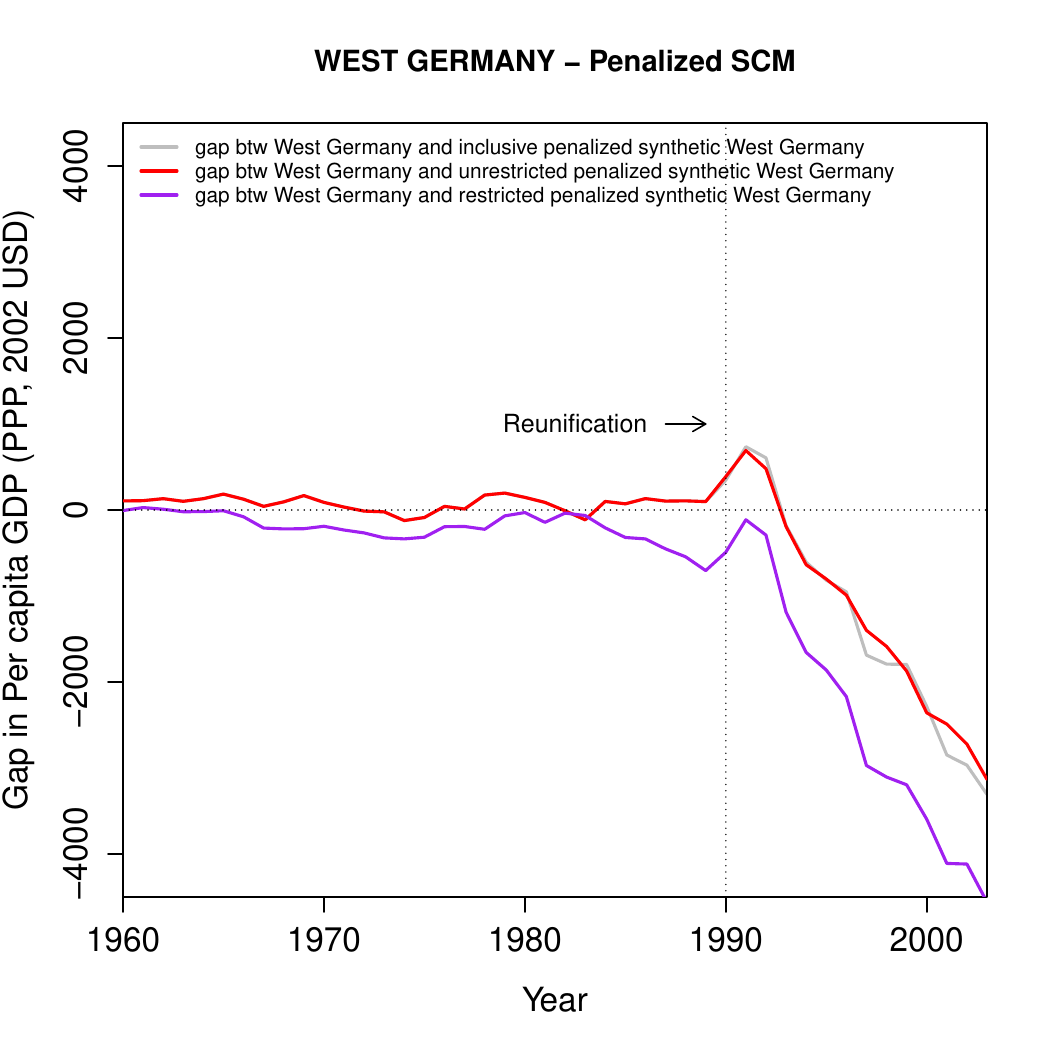} &
\includegraphics[width=0.46\textwidth]{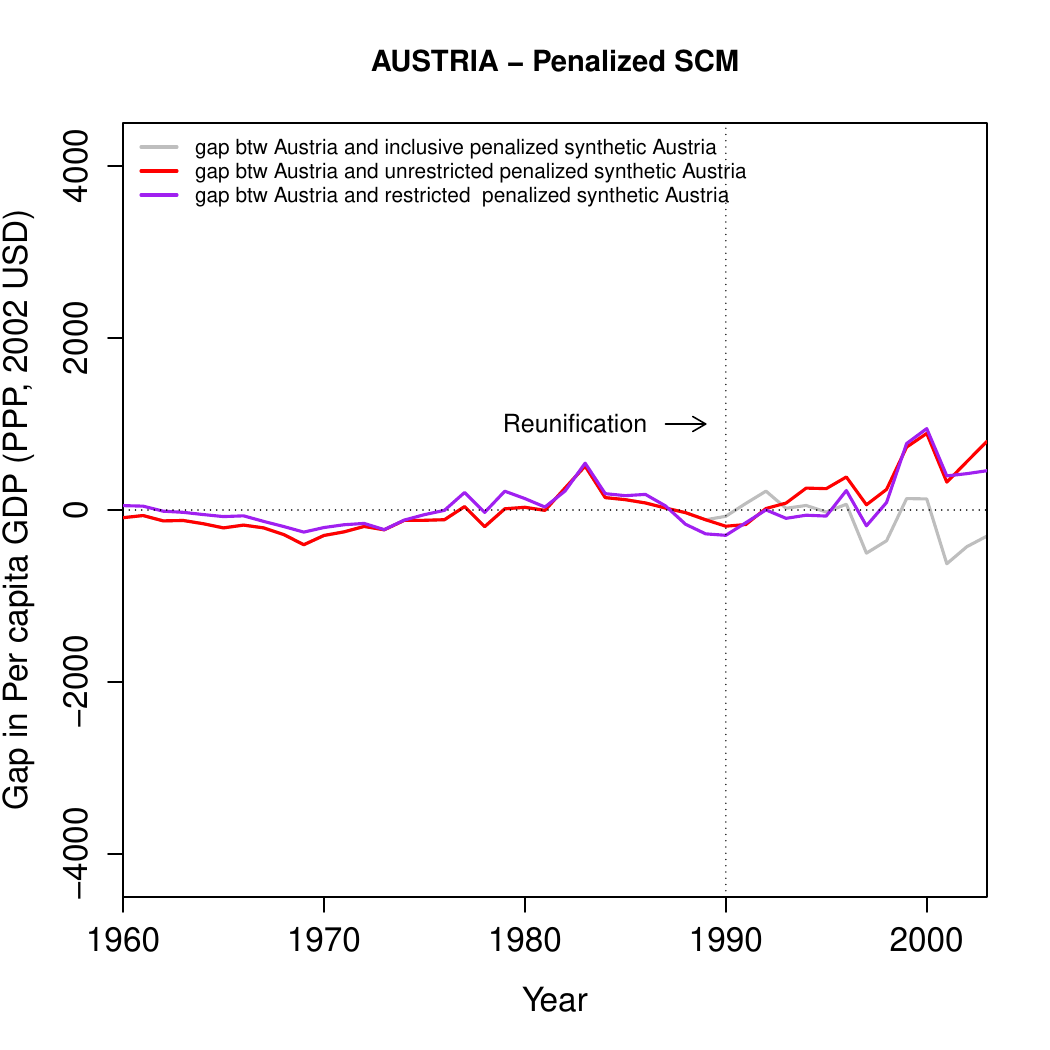} \\
{\footnotesize (a) Gaps, West Germany, penalized SCM} & {\footnotesize (b) Gaps, Austria, penalized SCM} \\[4pt]
\includegraphics[width=0.46\textwidth]{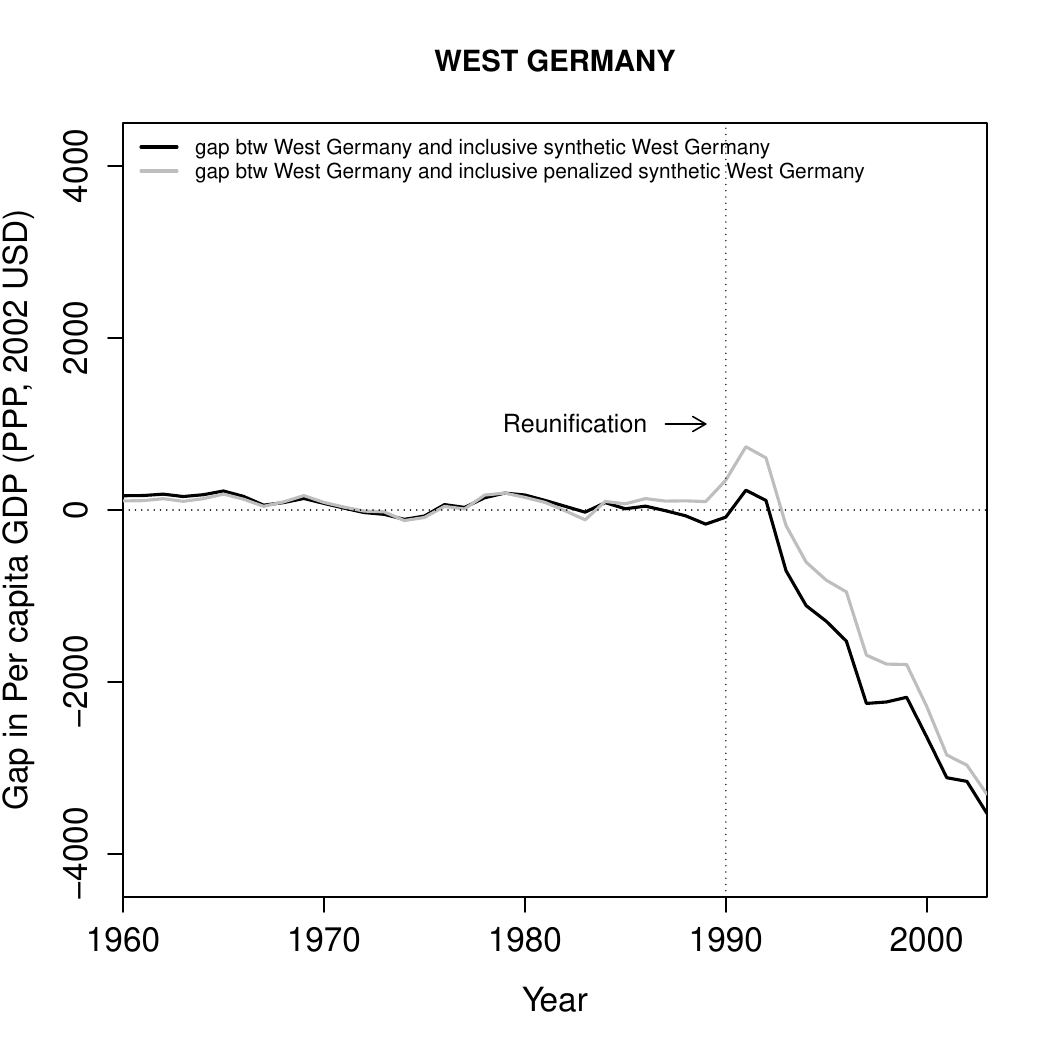} &
\includegraphics[width=0.46\textwidth]{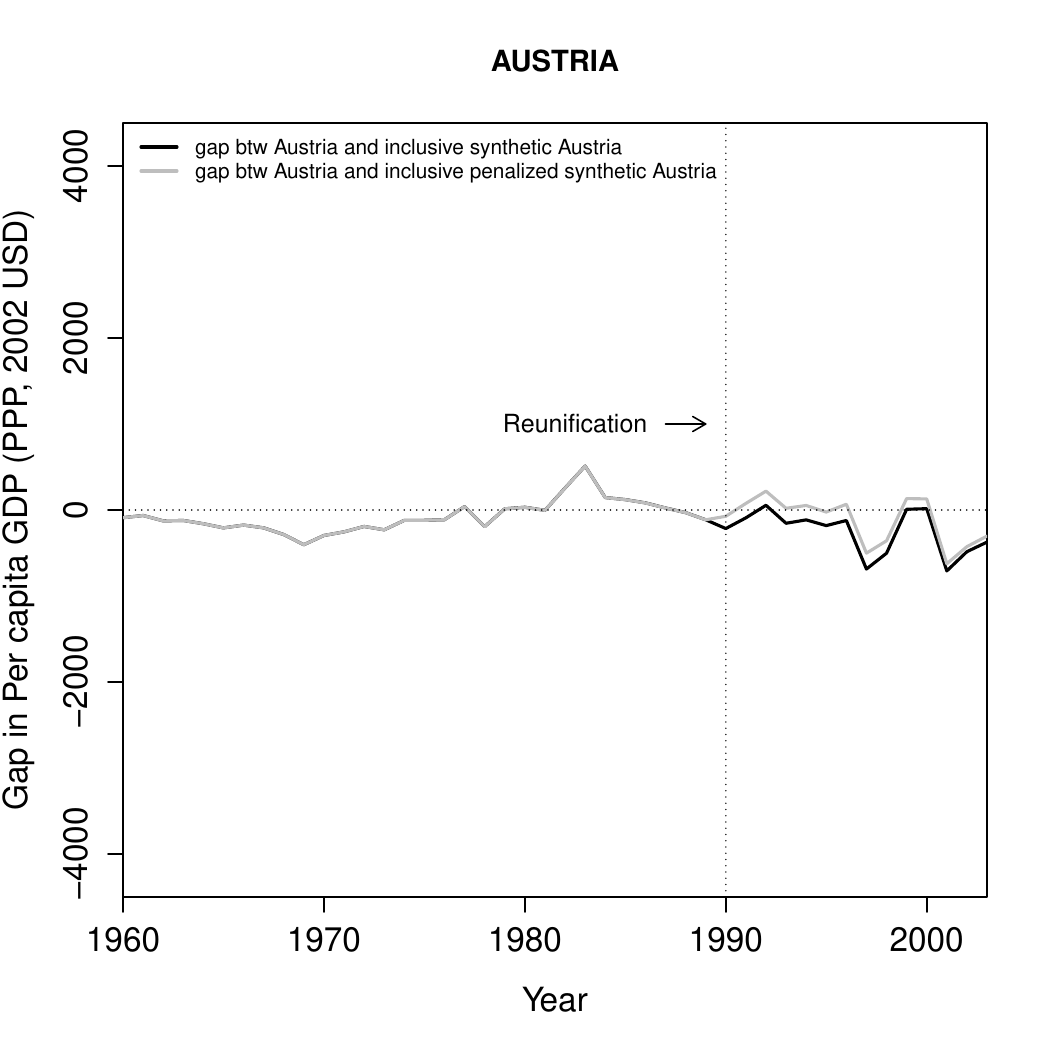} \\
{\footnotesize (c) Penalized iSCM vs iSCM, West Germany} & {\footnotesize (d) Penalized iSCM vs iSCM, Austria} \\
\end{tabular}
\caption{\label{fig:penalized}Estimated effects with the penalized SCM.}
\footnotesize
\renewcommand{\baselineskip}{11pt}
\textbf{Note:} Panels (a) and (b) plot, for West Germany and Austria respectively, the gaps between the observed GDP per capita and the corresponding synthetic control under the unrestricted, restricted and inclusive versions of the penalized SCM of \citet{Aba2019}, with the penalty parameter selected by cross-validation, over 1960--2003; the vertical line marks reunification in 1990. Panels (c) and (d) compare the gaps of the inclusive penalized SCM with those of iSCM based on the original SCM. The pre-intervention plotting convention of Figure~\ref{fig:main} applies.
\end{figure}

As iSCM can be combined with any SC-type estimator, we re-estimate the effects on West Germany and Austria using the penalized SCM of \citet{Aba2019}. We adopt the same specification as in the original SCM analysis, select the penalty parameter $\lambda$ by cross-validation and retain the same predictor weight matrix. The selected penalty is positive for West Germany ($\lambda \approx 0.05$) but essentially zero for Austria ($\lambda \approx 0$), so the penalized synthetic Austria is nearly identical to the original one. One might expect a method designed to improve the pre-intervention fit to mitigate the consequences of excluding Austria from the donor pool; the results suggest otherwise. Figure~\ref{fig:penalized}, panels (a) and (b), shows that the pre-intervention fit deteriorates substantially in the restricted penalized specification for both countries, especially for West Germany. For Austria, the selected penalty is close to zero, so the penalized SCM is nearly identical to the original SCM. Overall, the inclusive penalized SCM and iSCM yield very similar results, as shown in panels (c) and (d).

\section{Conclusion}\label{sec:conclusion}
\setcounter{equation}{0}

This paper has introduced iSCM, a procedure that allows units potentially affected by an intervention, directly or through spillovers, to remain in the donor pool of synthetic control methods. The procedure requires \textit{a priori} knowledge of which units may be affected, but it leaves the effects on those units fully unrestricted, and it can be implemented, at negligible computational cost, on top of the original synthetic control estimator or most of its recent extensions. The underlying intuition is that, because synthetic control weights are estimated on pre-intervention data only, each contaminated estimate is a known linear combination of the effects of interest, so that solving the resulting system removes the contamination exactly. What remains is the approximation error of the underlying fits, amplified by a factor that is directly computable from the estimated weights, and this factor anchors our practical guidance. When it is close to one, the full correction can be applied safely; when it is large, the simplified iSCM offers protection against the amplification at the cost of relying on restricted fits for the affected units; and when the restricted donor pool fits the main treated unit about as well as the unrestricted one, no correction is needed and the restricted SCM can be used directly. In all cases we recommend reporting all four specifications together with the amplification diagnostic.

In the German reunification application, keeping Austria in the donor pool and correcting for spillovers matters for both effects. The estimated effect of reunification on West Germany's GDP per capita becomes modestly more negative than the unrestricted SCM estimate, by about 7\% per year on average, whereas the restricted SCM, whose pre-intervention fit deteriorates sharply once Austria is excluded, produces an average estimate that is about 42\% larger in magnitude than the full iSCM estimate. For Austria, iSCM estimates a small negative spillover effect, of up to 709 US dollars of GDP per capita, consistent in sign with the conjecture of \citet{Aba2015}, although it cannot be distinguished from placebo variation; the unrestricted and restricted SCM instead deliver implausibly positive spillover estimates. The penalized SCM yields very similar conclusions.



\bibliographystyle{chicago}
\bibliography{main}

\section*{Appendix A: Proofs and supplementary results}
\renewcommand{\theequation}{A.\arabic{equation}}
\renewcommand{\thesection}{A}
\renewcommand{\thelemma}{A.\arabic{lemma}}
\setcounter{equation}{0}
\setcounter{lemma}{0}

\medskip

The following lemma provides easily verifiable sufficient conditions for the nonsingularity of $\widehat{\Omega}$ and for the M-matrix condition of Lemma~\ref{lem:purecontrol}.

\begin{lemma}\label{lem:mmatrix}
Suppose that, for each $i=1,\ldots,m$, the weights of the $i$-th SC-type fit are nonnegative and sum to one, and that the total weight assigned to the pure-control units is strictly positive, $\widehat{s}_{i}>0$. Then $\widehat{\Omega}$ is strictly diagonally dominant by rows and nonsingular. Moreover, $\widehat{\Omega}$ is a nonsingular M-matrix and $\widehat{\Omega}^{-1}\ge 0$ elementwise.
\end{lemma}

\noindent\textbf{Proof:} By construction, the diagonal entries of $\widehat{\Omega}$ equal one and the off-diagonal entries of row $i$ are the negatives of the weights that fit $i$ assigns to the other units in $\{1\}\cup M$. Under nonnegativity and adding-up, the sum of the absolute values of the off-diagonal entries of row $i$ equals the total weight assigned to those units, which is $1-\widehat{s}_{i}<1=\widehat{\Omega}_{ii}$. Hence $\widehat{\Omega}$ is strictly diagonally dominant by rows, and nonsingularity follows from the L\'evy--Desplanques theorem. Since the off-diagonal entries of $\widehat{\Omega}$ are nonpositive, $\widehat{\Omega}$ is a Z-matrix; a strictly diagonally dominant Z-matrix with positive diagonal entries is a nonsingular M-matrix, and the inverse of a nonsingular M-matrix is elementwise nonnegative; see \citet{BermanPlemmons1994}. \hfill$\square$

When some fit places zero total weight on the pure controls, $\widehat{\Omega}$ may be singular: for example, if $m=2$ and the main treated unit and the potentially affected unit assign weight one to each other, then $\widehat{w}_{2}=\widehat{l}_{1}=1$ and $\det(\widehat{\Omega})=0$. For SC-type estimators that allow for negative weights, diagonal dominance is not guaranteed, and the nonsingularity and conditioning of $\widehat{\Omega}$ should be checked directly, e.g.\ by computing $\sigma_{\min}(\widehat{\Omega})$.

\medskip

\noindent\textbf{Proof of Lemma~\ref{lem:purecontrol}:}
For each $i\in\{1,\ldots,m\}$, let $\widehat{q}_{ij}$, $j\in(\{1\}\cup M)\setminus\{i\}$, denote the estimated weights assigned to the main treated and potentially affected units in the $i$-th fit, so that the corresponding contaminated estimate can be written as
\begin{equation}
\widehat{\beta}_{it}=Y_{it}-\sum_{j\in(\{1\}\cup M)\setminus\{i\}}\widehat{q}_{ij}Y_{jt}-\widehat{p}_{i}'\,Y_{Ct}.
\end{equation}
By the construction of $\widehat{\Omega}$ in \eqref{eq:system}, stacking these $m$ equations yields the identity
\begin{equation}\label{eq:stacked}
\widehat{\beta}_{t}=\widehat{\Omega}\,(Y_{1t},Y_{2t},\ldots,Y_{mt})'-\widehat{P}\,Y_{Ct}.
\end{equation}
Premultiplying \eqref{eq:stacked} by $\widehat{\Omega}^{-1}$ gives
\[
\widehat{\vartheta}^{\,iSCM}_{t}=(Y_{1t},Y_{2t},\ldots,Y_{mt})'-\widehat{\Omega}^{-1}\widehat{P}\,Y_{Ct},
\]
and taking the first component yields $\widehat{\theta}^{\,iSCM}_{1t}=Y_{1t}-e_{1}'\widehat{\Omega}^{-1}\widehat{P}\,Y_{Ct}=Y_{1t}-Y_{Ct}'\,\widehat{a}$, which is \eqref{eq:purecontrolrep}.

For the adding-up property, note that under adding-up constraints the weights of fit $i$ on the units in $(\{1\}\cup M)\setminus\{i\}$ sum to $1-\widehat{s}_{i}$, so the $i$-th row sum of $\widehat{\Omega}$ equals $1-(1-\widehat{s}_{i})=\widehat{s}_{i}$, i.e.\ $\widehat{\Omega}\mathbf{1}_{m}=\widehat{s}$. Hence $\mathbf{1}_{m}=\widehat{\Omega}^{-1}\widehat{s}$, and since $\widehat{P}\mathbf{1}_{C}=\widehat{s}$,
\[
\mathbf{1}_{C}'\,\widehat{a}=e_{1}'\widehat{\Omega}^{-1}\widehat{P}\,\mathbf{1}_{C}=e_{1}'\widehat{\Omega}^{-1}\widehat{s}=e_{1}'\mathbf{1}_{m}=1.
\]
Finally, if $\widehat{P}\ge 0$ and $\widehat{\Omega}^{-1}\ge 0$, then $\widehat{a}=(e_{1}'\widehat{\Omega}^{-1}\widehat{P})'\ge 0$ elementwise.
\hfill$\square$

\noindent\textbf{Proof of Proposition~\ref{prop:implicit}:}
By \eqref{eq:tiscm},
\[
\widehat{\theta}^{\,iSCM}_{t}=\frac{\widehat{\theta}_{t}+\widehat{w}_{2}\widehat{\gamma}_{t}}{1-\widehat{w}_{2}\widehat{l}_{1}}.
\]
Substituting $\widehat{\theta}_{t}=Y_{1t}-\widehat{w}_{2}Y_{2t}-Y_{Ct}'\widehat{W}_{C}$ and $\widehat{\gamma}_{t}=Y_{2t}-\widehat{l}_{1}Y_{1t}-Y_{Ct}'\widehat{L}_{C}$ yields
\[
\widehat{\theta}^{\,iSCM}_{t}
=\frac{Y_{1t}\big(1-\widehat{w}_{2}\widehat{l}_{1}\big)-Y_{Ct}'\big(\widehat{W}_{C}+\widehat{w}_{2}\widehat{L}_{C}\big)}{1-\widehat{w}_{2}\widehat{l}_{1}}
=Y_{1t}-Y_{Ct}'\,\frac{\widehat{W}_{C}+\widehat{w}_{2}\widehat{L}_{C}}{1-\widehat{w}_{2}\widehat{l}_{1}},
\]
which is \eqref{eq:implicitrestricted}.
\hfill$\square$

\noindent\textbf{Proof of Corollary~\ref{cor:convexity}:}
Nonnegativity of $\widehat{W}^{\,*}_{C}$ is immediate from $\widehat{W}_{C}\ge 0$, $\widehat{L}_{C}\ge 0$, $\widehat{w}_{2}\ge 0$ and $\widehat{w}_{2}\widehat{l}_{1}<1$. For the adding-up property,
\[
\mathbf{1}_{C}'\widehat{W}^{\,*}_{C}
=\frac{(1-\widehat{w}_{2})+\widehat{w}_{2}(1-\widehat{l}_{1})}{1-\widehat{w}_{2}\widehat{l}_{1}}
=\frac{1-\widehat{w}_{2}\widehat{l}_{1}}{1-\widehat{w}_{2}\widehat{l}_{1}}=1.
\]
\hfill$\square$

\noindent\textbf{Proof of Corollary~\ref{cor:plugin}:}
Immediate from \eqref{eq:implicitrestricted} with $\widehat{l}_{1}=0$.
\hfill$\square$

\section*{Appendix B: Additional empirical results and data sources}
\renewcommand{\thesection}{B}
\renewcommand{\theequation}{B.\arabic{equation}}
\renewcommand{\thefigure}{B.\arabic{figure}}
\setcounter{equation}{0}
\setcounter{figure}{0}

\medskip

\begin{figure}
    \centering
    \includegraphics[width=1\linewidth]{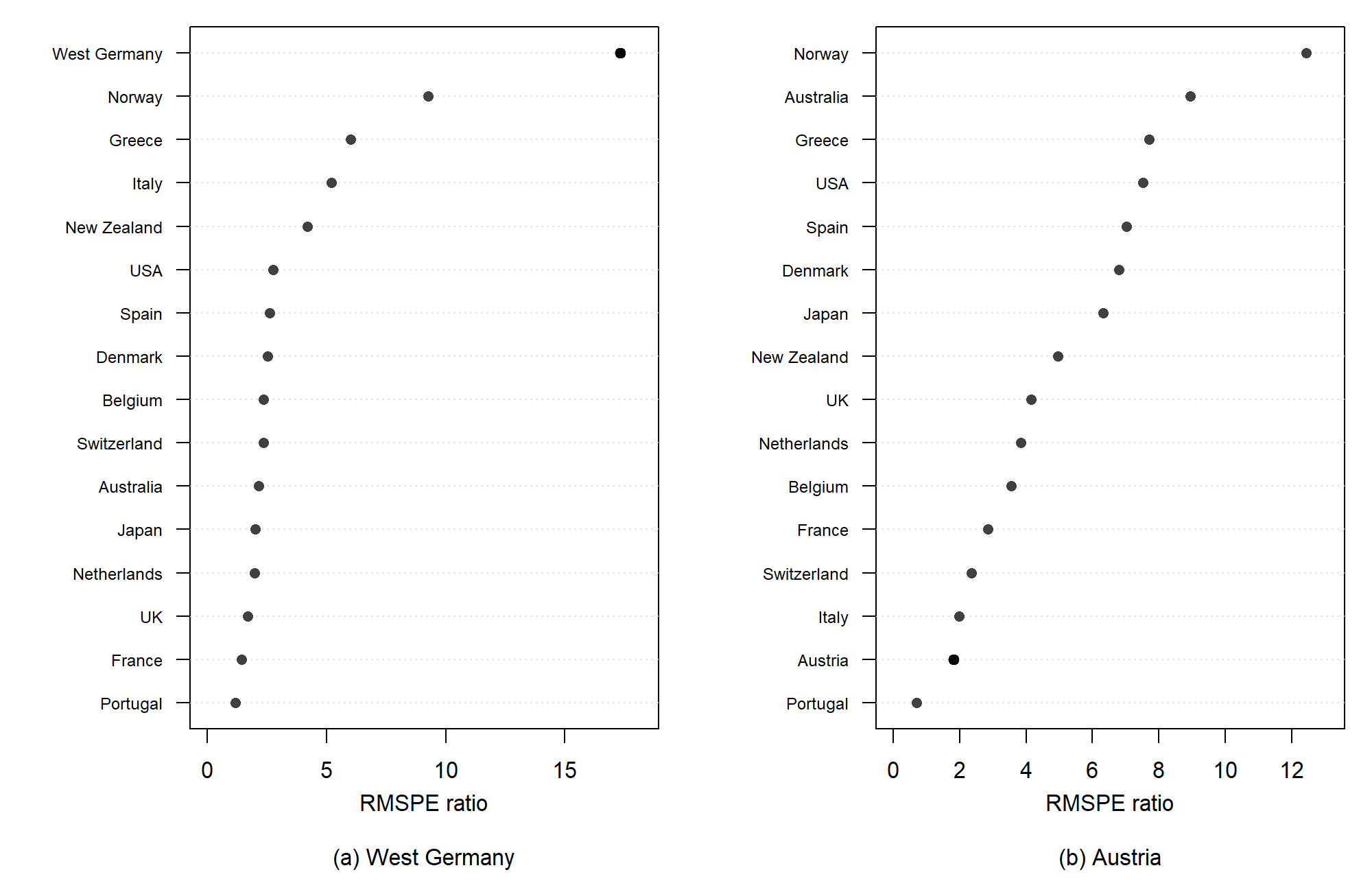}
    \caption{Post -- to pre-reunification RMSPE ratios}
    \label{fig:rmspe}
\end{figure}

Figure~\ref{fig:rmspe} shows, for the donor pools of West Germany and Austria, the ratios between the post- and pre-reunification RMSPEs used in the placebo analysis of Section~\ref{sec:inference}. West Germany's ratio is by far the largest in its donor pool, indicating an effect that is large relative to the placebo distribution. Austria's ratio is the second lowest in its donor pool, indicating that the estimated spillover effect cannot be distinguished from placebo variation.

\medskip

\noindent\textbf{Data sources.} We use the country-level panel data assembled by \citet{Aba2015} for 17 OECD countries over 1960--2003, available in the replication files of that paper. The outcome variable is real GDP per capita at purchasing power parity in 2002 US dollars, from the OECD National Accounts; the predictors are trade openness (exports plus imports over GDP), the inflation rate, the industry share of value added, the secondary-school attainment rate and the investment rate, from the OECD and the World Bank as documented by \citet{Aba2015}.

\section*{Appendix C: A geometric illustration}
\renewcommand{\thesection}{C}
\renewcommand{\theequation}{C.\arabic{equation}}
\renewcommand{\thefigure}{C.\arabic{figure}}
\setcounter{equation}{0}
\setcounter{figure}{0}

\medskip

\begin{figure}
\centering
{\includegraphics[width=0.6\textwidth]{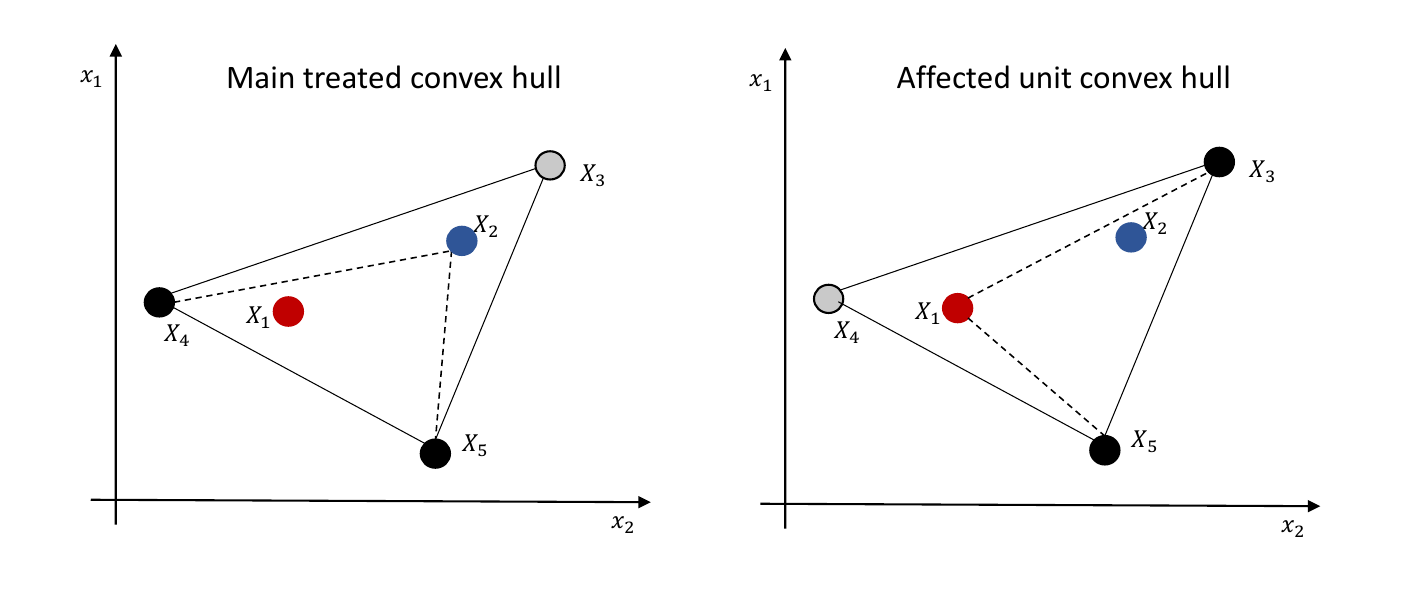}}
\caption{\label{fig:hull1}Both units inside the convex hull of the pure controls.}
\end{figure}

\begin{figure}
\centering
\begin{tabular}{c}
\includegraphics[width=0.6\textwidth]{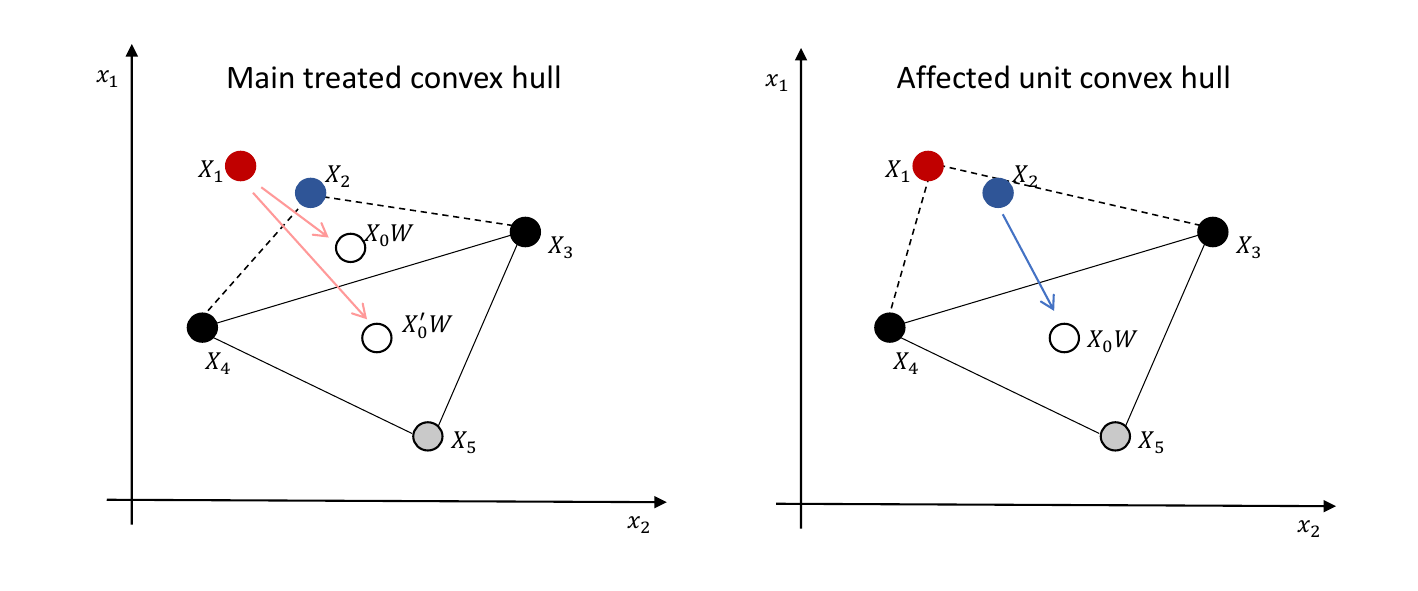}\\
{\footnotesize (a)}\\[2pt]
\includegraphics[width=0.6\textwidth]{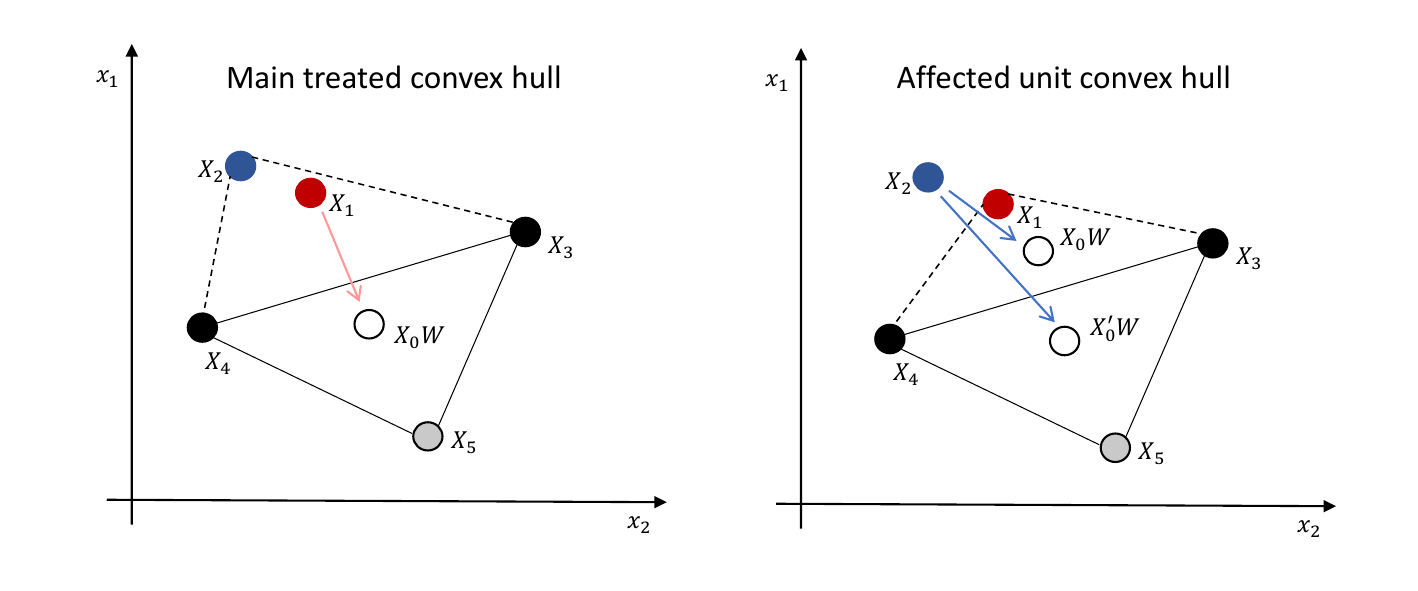}\\
{\footnotesize (b)}\\[2pt]
\includegraphics[width=0.6\textwidth]{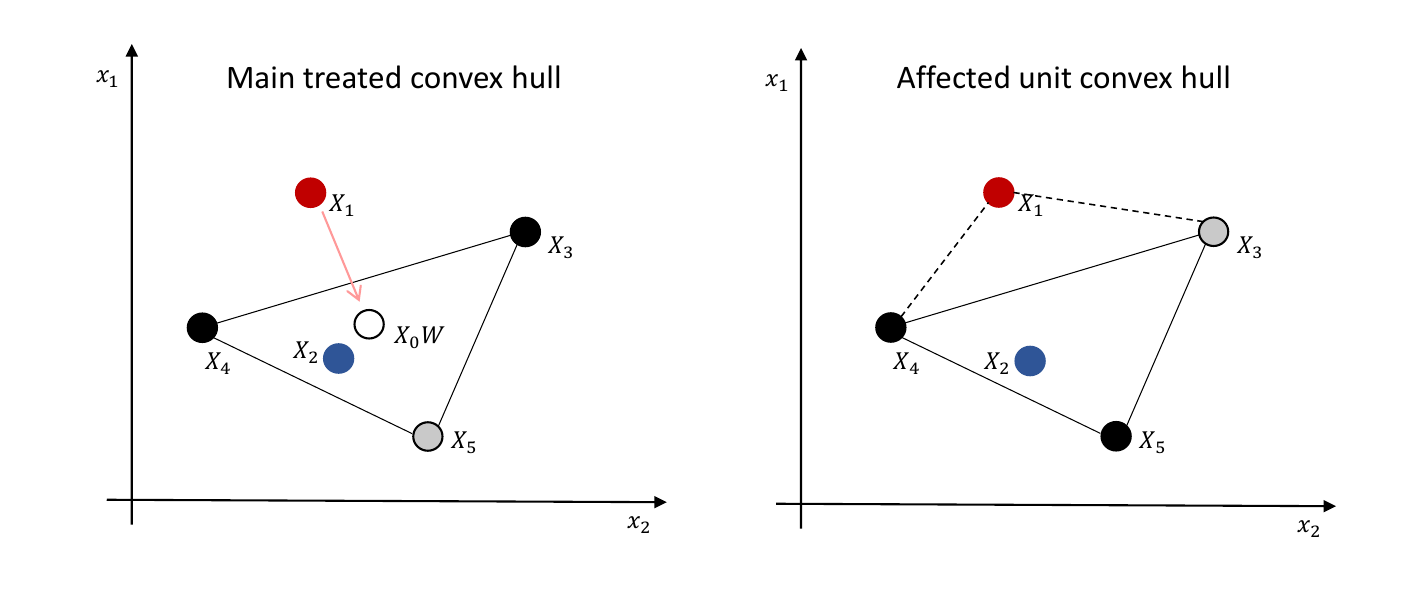}\\
{\footnotesize (c)}\\[2pt]
\includegraphics[width=0.6\textwidth]{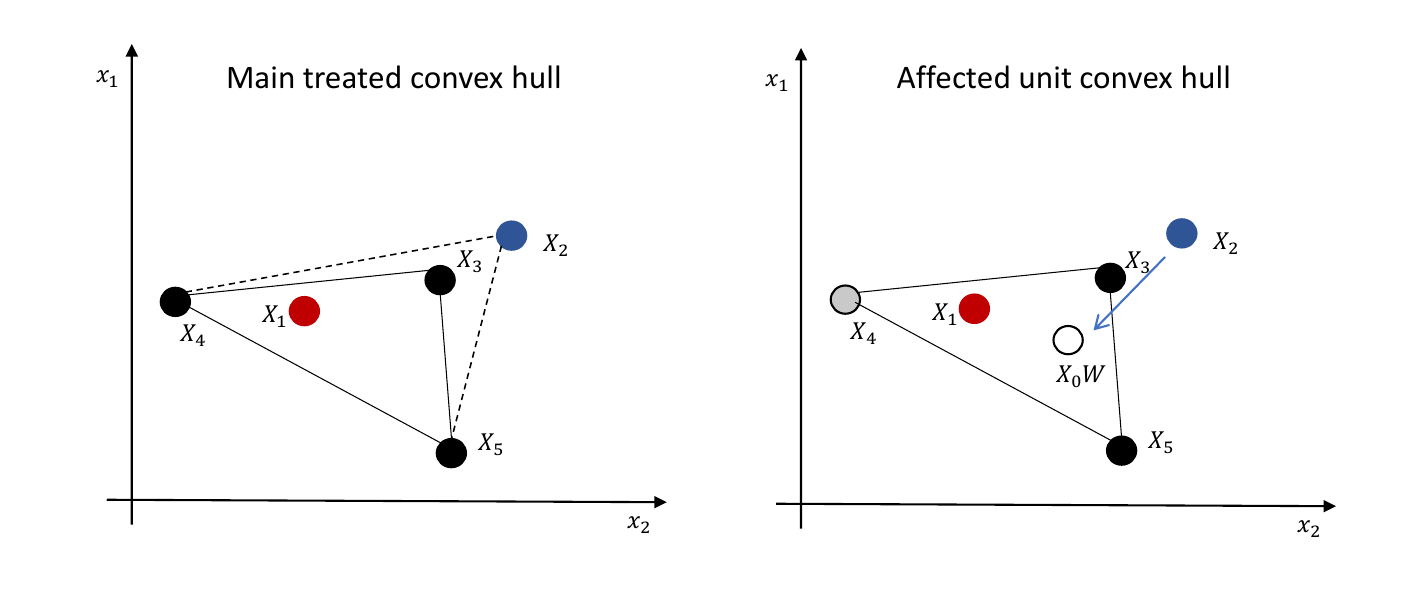}\\
{\footnotesize (d)}\\
\end{tabular}
\caption{\label{fig:hull2}At least one unit outside the convex hull of the pure controls.}
\end{figure}

This appendix illustrates geometrically when including potentially affected units in the donor pool is valuable, complementing the discussion in Section~\ref{sec:relationrestricted}. In Figures \ref{fig:hull1} and \ref{fig:hull2} we represent possible configurations from the point of view of the main treated unit (left side of each panel) and of the potentially affected unit (right side of each panel). Without loss of generality, each unit is characterized by two predictors, $x_{1}$ and $x_{2}$. The red point, $X_{1}$, collects the pre-intervention predictors of the main treated unit, and the blue point, $X_{2}$, those of the single potentially affected unit. All other points represent pure-control units: black points contribute to the relevant synthetic control, whereas grey points do not.

Figure~\ref{fig:hull1} shows the configuration in which both the main treated unit and the potentially affected unit lie \textit{inside the convex hull} of the pure controls. In this case $X_{1}$ can be reproduced using pure controls only. However, the right side of the panel shows that including the potentially affected unit in the donor pool allows the exclusion of the most distant pure control (unit 3), shrinking the convex hull used for interpolation and thereby potentially reducing interpolation bias (\citealp{Aba2020}). The same applies to the potentially affected unit: including the main treated unit allows the exclusion of unit 4, which is farthest from unit 2.

Figure~\ref{fig:hull2} presents configurations in which at least one of the two units lies \textit{outside the convex hull} of the remaining donors. In panel (a), the main treated unit lies outside but close to the hull: $X_{1}$ can only be approximated, but excluding unit 2 enlarges the discrepancy $X_{1}-X_{0}W$, so unit 2 should be included; on the right side of the panel, the potentially affected unit lies outside the hull unless the main treated unit is included in its donor pool. Panel (b) shows the symmetric situation. In panel (c), the main treated unit lies outside the hull and including unit 2 improves the approximation, while unit 2 always lies inside the hull, and including unit 1 allows the exclusion of unit 3, improving the approximation. Panel (d) describes a configuration in which using iSCM is clearly counterproductive; in this configuration, however, the potentially affected unit would receive little or no weight, so the recommendation of Section~\ref{sec:guidance} to apply iSCM only when potentially affected units receive substantial weight makes this case unlikely to be relevant in practice.

\end{document}